\documentclass[conference]{IEEEtran}
\usepackage{tikz}
\usepackage{url}
\usepackage{cite}
\usepackage{amsmath,amssymb,amsfonts}
\usepackage{graphicx}
\usepackage{textcomp}
\usepackage{xcolor}
\usepackage{subcaption}
\usepackage{multirow}
\usepackage{pifont}
\usepackage{booktabs}
\usepackage{siunitx}

\usepackage{soul}

\usepackage{breakurl}

\usepackage[colorlinks]{hyperref}
\setstcolor{red}

\usepackage{algorithm, algorithmicx}
\usepackage{algpseudocode}

\let\oldReturn\Return
\renewcommand{\Return}{\State\oldReturn}
\algnewcommand\algorithmicforeach{\textbf{for each}}
\algdef{S}[FOR]{ForEach}[1]{\algorithmicforeach\ #1\ \algorithmicdo}

\newlength{\bibitemsep}
\newlength{\bibparskip}
\let\oldthebibliography\thebibliography
\renewcommand\thebibliography[1]{%
  \oldthebibliography{#1}%
  \setlength{\parskip}{\bibitemsep}%
  \setlength{\itemsep}{\bibparskip}%
}

\iftrue

\newcommand{\cut}[1]{}
\newcommand{\alfred}[1]{}
\newcommand{\junjie}[1]{}
\newcommand{\ziwen}[1]{}
\newcommand{\yunpeng}[1]{}

\newcommand{\todo}[1]{}

\newcommand{\ld}{$\mathrm{LD^3}$}
\newcommand{\ldi}{$LD^3$}
\newcommand{\fr}{\textit{FusionRipper}}

\newcommand{\newparts}[1]{#1}

\else

\newcommand{\cut}[1]{}
\newcommand{\alfred}[1]{\textcolor{red}{[Alfred: #1]}}
\newcommand{\junjie}[1]{\textcolor{orange}{[Junjie: #1]}}
\newcommand{\ziwen}[1]{\textcolor{red}{[Ziwen: #1]}}
\newcommand{\yunpeng}[1]{\textcolor{red}{[Yunpeng: #1]}}

\newcommand{\todo}[1]{\textcolor{red}{[TODO: #1]}}

\newcommand{\ld}{$\mathrm{LD^3}$}
\newcommand{\ldi}{$LD^3$}
\newcommand{\fr}{\textit{FusionRipper}}

\newcommand{\newparts}[1]{{\color{blue}#1}}

\fi

\newcommand{\diffst}[1]{}

\makeatletter
\renewcommand\p@subfigure{\thefigure\,}

\makeatother

\usepackage{mdframed}
\mdfdefinestyle{rebuttalstyle}{innerleftmargin=3pt, innerrightmargin=3pt, innertopmargin=3pt, innerbottommargin=3pt}

\newcounter{response}[section]

\newcounter{revision}[section]

\newcounter{comment}[section]

\iftrue

\newcommand{\nsection}[1]{\section{#1}}
\newcommand{\nsubsection}[1]{\subsection{#1}}

\else

\newcommand{\nsection}[1]{\vspace{-0.32cm}\section{#1}\vspace{-0.25cm}}
\newcommand{\nsubsection}[1]{\vspace{-0.32cm}\subsection{#1}\vspace{-0.25cm}}

\fi

\begin{document}
%-------------------------------------------------------------------------------

\title{\Large \bf Lateral-Direction Localization Attack in High-Level Autonomous Driving:\\Domain-Specific Defense Opportunity via Lane Detection}

%\author{\IEEEauthorblockN{{\rm Junjie Shen}$^{}$\quad {\rm Yunpeng Luo}$^{}$ \quad {\rm Ziwen Wan}$^{}$ \quad {\rm Qi Alfred Chen}}
%\IEEEauthorblockA{$^{}$University of California, Irvine}
%\IEEEauthorblockA{
%$^{}$\{junjies1, yunpel3, ziwenw8, alfchen\}@uci.edu}}

\author{{\rm Junjie Shen}$^{}$\quad {\rm Yunpeng Luo}$^{}$ \quad {\rm Ziwen Wan}$^{}$ \quad {\rm Qi Alfred Chen}\\% <-this % stops a space
University of California, Irvine \ \ \ \{junjies1, yunpel3, ziwenw8, alfchen\}@uci.edu}

\maketitle

\begin{abstract}
Localization in high-level Autonomous Driving (AD) systems is highly security critical. While the popular Multi-Sensor Fusion (MSF) based design can be more robust against single-source sensor spoofing attacks, it is found recently that state-of-the-art MSF algorithms is vulnerable to GPS spoofing alone due to practical factors, which can cause various road hazards such as driving off road or onto the wrong way. In this work, we perform the first systematic exploration of the novel usage of lane detection (LD) to defend against such attacks. We first systematically analyze the potentials of such a domain-specific defense opportunity, and then design a novel LD-based defense approach, \ld{}, that aims at not only detecting such attacks effectively in the real time, but also safely stopping the victim in the ego lane upon detection considering the absence of onboard human drivers.

We evaluate \ld{} on real-world sensor traces and find that it can achieve effective and timely detection against existing attack with 100\% true positive rates and 0\% false positive rates. Results also show that \ld{} is robust to diverse environmental conditions and is effective at steering the AD vehicle to safely stop within the current traffic lane. We implement \ld{} on two open-source high-level AD systems, Baidu Apollo and Autoware, and validate its defense capability in both simulation and the physical world in end-to-end driving. We further conduct adaptive attack evaluations and find that \ld{} is effective at bounding the deviations from reaching the attack goals in stealthy attacks and is robust to latest LD-side attack.

%that tries to disrupt the attack response stage.

%As discovered in this work, the recent physical-invariant based defenses fail to detect such attack in AD context , and the existing attack recovery methods cannot apply due to the same issue. Thus, to solve this, we identify a defense-suitable information source in high-level AD systems and design a systematic defense solution with both attack detection and attack response stages to detect such lateral-direction localization attack and safety drive the vehicle to stop in the ego lane after attack detection.

%High-level Autonomous Driving (AD) systems predominantly adopt a Multi-Sensor Fusion (MSF) based localization design to achieve centimeter-level accuracy as required for making safety-critical driving decisions. But a recent attack is able to use GPS spoofing alone to cause large lateral deviations in the MSF outputs, which can lead the AD vehicle to drive off road or onto the wrong way. However, as discovered in this work, the recent physical-invariant based defenses fail to detect such attack due to inaccurate state estimation models, and the existing attack recovery methods cannot apply due to the same issue. Thus, to solve this, we identify a defense-suitable information source in high-level AD systems and design a systematic defense solution with both attack detection and attack response stages to detect such lateral-direction localization attack and safety drive the vehicle to stop in the ego lane after attack detection.

\end{abstract}

% 06/01/21: adjust the vspace after all figure/table/alg
% \setlength{\textfloatsep}{10pt}
% \setlength{\floatsep}{10pt}
\setlength{\textfloatsep}{6pt}
\setlength{\floatsep}{6pt}

\nsection{Introduction} \label{sec:intro}

Recently, high-level Autonomous Driving (AD) vehicles~\cite{sae2021}, e.g., Level-4 ones, are gradually becoming part of the transportation system by providing commercial services such as self-driving taxis~\cite{baidu_driverless_robotaxi, waymo-one}, buses~\cite{baidu_bus_service, uk_bus}, and trucks~\cite{tusimple-truck, aurora_truck}. In particular, AD companies such as Waymo and Baidu are already offering commercial RoboTaxi services without safety drivers~\cite{baidu_driverless_robotaxi, waymo_driverless}, and more others are performing tests on public roads~\cite{cruise_driverless_testing, pony_driverless_testing}. To achieve high driving automation, the \textit{high-level AD system} (the ``\textit{brain}'') in such a vehicle needs to localize itself with \textit{centimeter-level} accuracy on the map~\cite{levinson2007map, reid2019localization, ega_requirement_report} to ensure safe and correct driving. Thus, today's industry-grade high-level AD systems predominantly adopt a Multi-Sensor Fusion (MSF) based localization design, which combines sensor inputs, typically GPS, LiDAR, and IMU, for overall higher accuracy and robustness in practice~\cite{wan2018robust, gao2015ins, soloviev2008tight, udacity_av_apollo, udacity_av_nd, coursera_av}. 

Due to the reliance on sensor inputs, AD localization is inherently vulnerable to sensor spoofing attacks, in particular GPS spoofing~\cite{fusionripper, spoof_tesla}, a long-existing security problem that is fundamentally difficult in both prevention and detection in practice~\cite{psiaki2016gnss, fusionripper}. Although the MSF-based design is generally more robust against such single-source sensor attacks, recent work~\cite{fusionripper} find that state-of-the-art MSF algorithms are still vulnerable to strategic GPS spoofing attacks due to non-deterministic and practical factors such as sensor noises and algorithm inaccuracies. To leverage such non-deterministic vulnerabilities, the authors devise a lateral-direction localization attack named \fr{} to \textit{opportunistically} inject lateral deviations in the MSF localization outputs, which will be translated into lateral deviations in the physical world by the AD control. Such lateral-direction localization attack is especially safety-critical in the AD context due to the potential consequences of road departure~\cite{road_depart_safety}.

% Specifically, the attacker would keep tailgating a victim AD vehicle until a vulnerable period occur
% and waits for a vulnerable period to inject large lateral deviations to cause the victim to drive off road or onto a wrong way.

% that as long as the attacker can tailgate a victim AD vehicle for certain duration (e.g., 2 min), GPS spoofing alone can take-over the fusion process and cause the victim to drive off road or onto a wrong way with high attack success rates.

%a wide range of road hazards such as driving off road or onto the wrong way even if the perception module is functioning perfectly~\cite{fusionripper}.

%\alfred{why emphasize ``high-level'' here?} \junjie{I want to distinguish them with low-level ones since they don't need centimeter level accuracy.}

%Although MSF is generally believed to have the potential to defeat sensor attacks such as GPS spoofing~\cite{davidson2016controlling, lee2017attack, zeng2018all, cardenas2019cyber}, recent work~\cite{fusionripper} discovers a fundamental design-level vulnerability in the representative MSF localization algorithms in academia and industry that allows attacker to use GPS spoofing alone to cause large \textit{lateral deviations} in MSF outputs, i.e., deviating to left or right. Leveraging this, they design an attack named \fr{}, which can deviate the AD vehicle to drive off road or onto a wrong way with high attack success rates.

%may be potentially applicable to high-level AD systems

So far, no software-based defenses have been proposed to defend against such latest lateral-direction localization attack in high-level AD systems. The closed-related ones are the recent \textit{physical-invariant} based defenses for small robotic vehicles such as drones and rovers~\cite{savior, ci}. Such defenses estimate system states (i.e., vehicle positions) based on control commands and use them to validate GPS signals.
% that uses physical-invariants based defenses~\cite{ci, savior}, where they estimate system states (i.e., vehicle positions) based on the control commands and use them to validate the GPS signals.
% To defend against such attack, one possibility is a recently-proposed defense direction that applies physical-invariants to estimate system states (e.g., vehicle positions) and use them to cross-check the sensor inputs (e.g., GPS position)~\cite{ci, savior}. 
While these works show high effectiveness for small robotic vehicles, we find that they have limited effectiveness in the AD context (evaluated as a baseline in~\S\ref{sec:eval_detection}), because (1) vehicle driving motions in the real-world are more diverse and complex (e.g., commonly have high-speed or curvy-road driving), and thus harder to model accurately, and (2) attack deviation goals can be much smaller while still being safety-critical, e.g., even less than 0.5-meter lateral deviations can cause lane departure. 
%with obvious attack deviation goals (e.g., $\sim$50 meters~\cite{savior})
%However, in this work, we find that the existing physical-invariants used in these works, e.g., bicycle model~\cite{kong2015kinematic}, are not accurate enough to detect stealthy attacks such as \fr{} (\S\ref{sec:motivation_savior}).
Moreover, these works did not consider the \textit{attack response} step after detection, which is especially critical for high-level AD systems due to the complex driving environment and the absence of onboard human drivers. A few recent works considered such attack response designs, e.g., attack recovery upon attack detection~\cite{choi2020software, zhang2020real}. However, they rely on similar state estimation models as above to replace the attacked sensors during the recovery period, which thus suffer from the same motion model accuracy limitations in AD context, and also counted on human operators to take over as soon as possible since such state estimations cannot replace physical sensors for a prolonged duration due to drifting~\cite{choi2020software}. Last but not least, they assume an effective attack detection in place, which does not yet exist in for high-level AD localization.

%the assumption that a human operator can exist to be ready to take over upon attack detection. 
%emergency operators. 
%Moreover, existing attack recovery works~\cite{choi2020software, zhang2020real} also suffer from the same problem as they also rely on similar state estimation models as replacement for the attacked sensors during the recovery period (typically 3--7 sec~\cite{eriksson2017takeover, gold2013take}) before taken over by the emergency operator, which high-level AD vehicles do not possess.

Compared to small robotic vehicles, the AD context may also have its unique defense opportunity for lateral-direction localization attacks, for example \textit{Lane Detection (LD)}~\cite{hillel2014recent, pan2018spatial}, which is directly related to the attack goals since it can measure the vehicle's physical lateral deviation in the ego lane in the real time. Today, LD is already widely used in low-level AD localization (e.g., for automated lane centering). However, due to its fundamental limit in achieving effective global localization (\S\ref{sec:opportunity}), it is currently \textit{not} used for high-level AD localization purposes. While less suitable for accuracy purposes, so far no prior works have explored its potential for \textit{defense purposes} in high-level AD localization.

%\textit{However, no prior works have explored LD for defense purposes beyond its original usages, e.g., automated lane centering in low-level AD systems.}

In this work, we thus perform the first concrete exploration of LD as a domain-specific defense opportunity for lateral-direction localization attacks in high-level AD systems. We start by systematically analyzing its high-level defense potential,
%qualitatively and quantitatively, 
and find that such an LD-based defense strategy has various design-level benefits such as generality to lateral-direction attacks, technology maturity, direct deployability, and independence to existing attacks. One potential downside is the lack of defense capability when lane line markings are not available (e.g., in intersections), but since latest lateral-direction attack design is fundamentally opportunistic, the attacker \textit{cannot} deterministically control the triggering of a desired deviation only at road regions without lane line markings. In fact, we find that a defense coverage of the road regions with lane line marking can already provide protection for \textit{99.2\%} of such opportunistic attack attempts (\S\ref{sec:opportunity}).

\begin{figure}[tbp]
\centering
\includegraphics[width=\columnwidth]{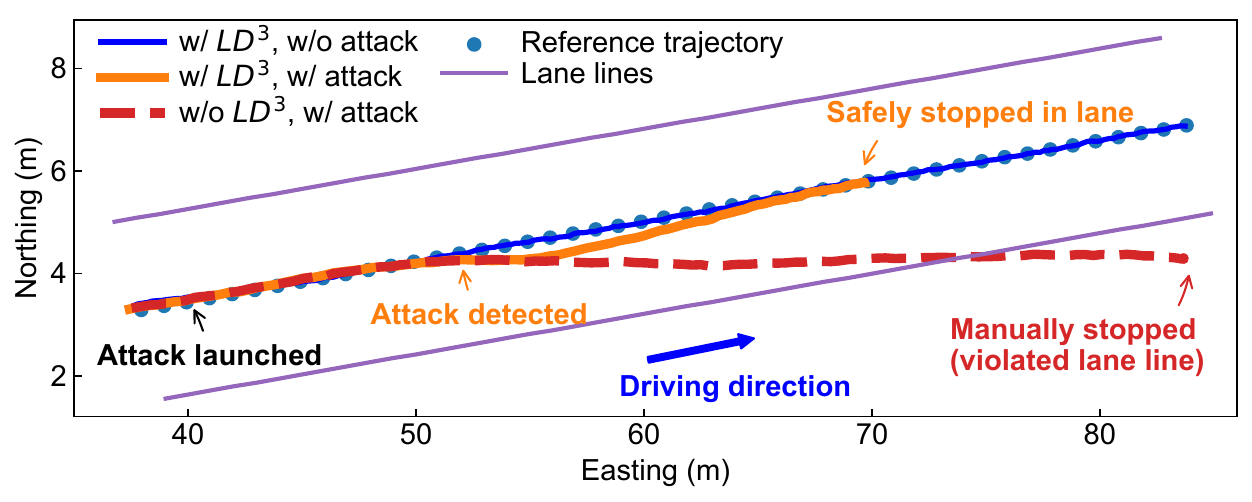}
\includegraphics[width=\columnwidth]{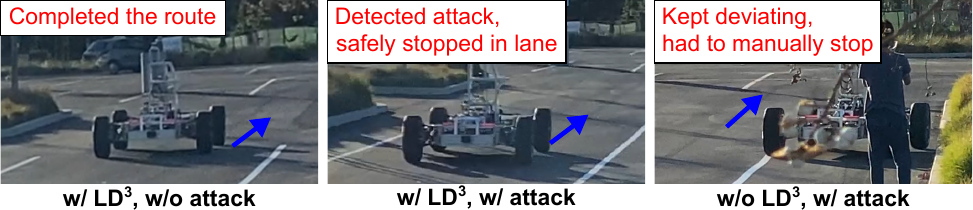}
% \vspace{-0.25in}
\caption{Physical-world end-to-end demonstrations of \ld{} effectiveness using a Level-4 AD development chassis of real vehicle size and with full closed-loop control. (Top) Vehicle driving trajectories in bird's eye view. (Bottom) Final stopping positions under the three experimental settings. The driving direction and vehicle heading are annotated with blue arrows.}
\label{fig:pixkit_stop_positions}
\vspace{-0.05in}
\end{figure}

Motivated by such multi-dimensional defense potentials, we design the first domain-specific LD-based defense approach called \ld{} (\underline{L}ane \underline{D}etection based \underline{L}ateral-\underline{D}irection \underline{L}ocalization attack \underline{D}efense), which is capable of both real-time attack \textit{detection} and \textit{response}. To use LD for attack detection, a tradeoff is that at which information level (i.e., GPS or MSF output) should we perform the detection. Recognizing that GPS outputs naturally have large noises and existing attack cannot deterministically predict when and where will large deviations occur in MSF, we decide to detect at the MSF output level to take advantage of such attack non-determinism.
%Since LD and MSF measure the vehicle's position in different coordinate systems (i.e., local vs global), we first convert the MSF outputs to lateral deviations to the lane centerline to make it comparable to the LD measurements. Specifically, a CUmulative SUM (CUSUM) anomaly detector is used to check the agreement between MSF and LD for attack detection (\S\ref{sec:design_detection}).
In the attack response (AR) stage, we choose to safely stop the vehicle in the ego lane, since this can minimize the attackable duration after detection and thus fundamentally bounds the attack-achievable deviation in the AR period. To account for the inherent LD-side adaptive attack surface introduced by \ld{}, we further design a novel safety-driven fusion between LD and MSF that systematically penalizes the source that is more aggressive in causing lateral deviations, which can fundamentally reduce the attacker's capability in causing safety damages in AR period even in adaptive settings. 

%. As AD vehicles may operate at high speeds, a reliable localization is necessary to steer the vehicle during the safe stopping, which may last for $\sim$7 seconds when on the highway (\S\ref{sec:eval_ar}). Since either MSF or LD could be attacked, we design a \textit{safety-driven fusion} to calculate a fused localization for the response. In particular, the fusion design penalizes the more aggressive source among MSF and LD in the AD driving context, to prevent the fused localization from being biased towards the attacked one (\S\ref{sec:design_ar}). With the fused localization, the AD planning and control then enforce an attack response trajectory, which follows the lane centerline, to safely stop the vehicle within the current lane.

We evaluate our defense against the latest lateral-direction localization attack on a diverse set of real-world sensor traces with various environmental conditions. 
% from the KAIST dataset~\cite{jeong2019complex} and also from driving our own vehicle with ADAS localization capabilities (global positioning and LD). 
Our results show that \ld{} is much more effective at detecting the attack compared to direct adaptation of physical-invariant based detection for small robotic vehicles~\cite{savior}. Specifically, \ld{} can achieve effective detection with 100\% true positive rates and 0\% false positive rates on the sensor traces and the detections are timely when the lateral deviations are not yet large enough to touch the lane boundaries. Moreover, \ld{} is also effective at keeping the AD vehicle within the lane during the attack response periods, where the vehicle's final stopping deviations are \textit{always} smaller than the lane straddling deviation. We also collect a night-time driving trace and find that \ld{} also has high defense robustness in low visibility conditions. 

%Our results show that \ld{} can achieve similar results as in the other traces, indicating a high defense robustness.

To further evaluate the defense in end-to-end driving with closed-loop control, we implement \ld{} on two open-source high-level AD systems, Baidu Apollo~\cite{apollo} and Autoware~\cite{autoware}, and evaluate in an industrial-grade AD simulator~\cite{lgsvl} and the physical world with a real vehicle-sized AD chassis. 
Our results show consistent results in end-to-end drivings as in the trace-based evaluations. Fig.~\ref{fig:pixkit_stop_positions} shows the vehicle driving trajectories and stopping positions in the physical world experiments. As shown, \ld{} can promptly detect the attack and safely stop the vehicle at the center of the lane, while without \ld{}, the vehicle drives out of lane boundary, and we have to manually stop the vehicle to prevent the collision.
The demo videos of the simulation and physical world experiments are available at
\textbf{\url{https://sites.google.com/view/cav-sec/LD3}}.

Lastly, we explore two potential adaptive attacks against \ld{}: (1) an ideal stealthy attack with the full knowledge of the defense aiming to evade the detection, and (2) the latest LD-side attack~\cite{sato2021dirty} against production AD systems. Results show that \ld{} can effectively bound the deviations of the stealthy attack from reaching the attack goals and can safely stop vehicle under the LD-side attack.

%\alfred{talk about stealthy attack?} \junjie{added, marked in red above.}

% \junjie{possible arguement: GPS is also not accurate, then why MSF use it? --> GPS has the benefit of not requiring map information, and not subject to map inaccuracy or outdateness --> good availability. But LiDAR and camera both requires maps; LiDAR is much more accurate and less environmentally restricted than camera. Thus MSF uses GPS \& LiDAR at this point.}

%Since prior works~\cite{kang2020lane, evlampev2020map} have explored global positioning methods using lane boundaries information, thus an alternative design is to use the LD-based global position as an additional fusion input for MSF localization to improve its robustness against lateral-direction attack. However, it is likely that adding such fusion input would degrade the MSF accuracy since their methods can only achieve 0.5m-level accuracy. In addition, such design cannot fundamentally prevent \fr{} since the vulnerability is due to sensor noises and algorithm inaccuracies~\cite{fusionripper}. In contrast, \ld{} is able to achieve perfect attack detection and can also steer the vehicle to safely stop within the ego lane after detection.

In summary, this work makes the following contributions:
\vspace{-\topsep}
\begin{itemize}
\setlength{\itemsep}{0pt}
\setlength{\parskip}{0pt}
    \item We perform the first systematic exploration of using LD to defend against lateral-direction localization attacks on high-level AD systems. We quantitatively show that LD can provide comprehensive defense coverage for existing attacks despite the reliance on lane line markings, and is independent of the AD localization inputs.
    \item We design \ld{}, a real-time defense solution including both attack detection and response stages. Evaluation on real-world sensor traces shows that \ld{} can achieve effective and timely attack detection, and can effectively stop the vehicle safely within the current lane. We also validate the robustness of \ld{} under low visibility conditions on a night-time driving trace.
    \item We implement \ld{} on two popular open-source AD systems, Baidu Apollo and Autoware, and evaluate the defense in end-to-end drivings in both simulation and the physical world.
    \item We evaluate \ld{} against two adaptive attacks and show that it is effective at bounding the deviations in the stealthy attack from reaching the attack goals and is robust to recent LD-side attack.
    % evaluate \ld{} on real-world sensor traces and end-to-end simulation environments. Results show that \ld{} can achieve perfect and timely attack detection, and can effectively steer the vehicle to stop within the ego lane. 
    % We also find that the timing overhead of \ld{} is negligible via evaluations on an embedded ADAS device.
\vspace{-\topsep}
\end{itemize}

% \item We evaluate the recent physical-invariant based defenses and find that they have limited effectiveness in the AD context due to inaccurate state estimation models.
% \vspace{-0.05in}
\nsection{Background and Threat Model} \label{sec:background}
% \vspace{0.1in}

\nsubsection{High-level AD Localization and MSF} \label{sec:background_msf}
\vspace{0.05in}

Today's high-level (e.g., Level-4~\cite{sae2021}) AD systems widely adopt a modular design with functional components such as localization, perception, prediction, planning, and control~\cite{apollo, autoware, udacity_av_apollo, udacity_av_nd, coursera_av}. Among them, localization is one of the most important modules that provides global positioning on the map for other modules such as planning and control to make safety-critical driving decisions. Since high-level AD systems need to navigate on the roads complete autonomously without any drivers, a localization with \textit{centimeter-level} accuracy is required to localize the AD vehicle on the traffic lane~\cite{levinson2007map, reid2019localization, ega_requirement_report}. 
High-level AD systems are typically equipped with various positioning sensors with diverse properties. For example, GPS provides global positioning with high availability, however, it often contains large positioning noises due to satellite signal transmission interferences and multi-path effect~\cite{gps_error_sources}; on the other hand, LiDAR localization algorithms (LiDAR locators) are able to accurately position the vehicle on a prebuilt LiDAR reflectance map using point cloud matching~\cite{wan2018robust, ndt, gao2015ins, levinson2010robust}. However, LiDAR locator performance can be severely degraded under adverse weather conditions or with an outdated LiDAR map. Thus, to achieve both high accuracy and robustness, high-level AD systems predominantly adopt a \textit{Multi-Sensor Fusion} (MSF) based localization design to \textit{leverage the strengths and compensate the weaknesses of different sensors} such as GPS, LiDAR, and IMU~\cite{wan2018robust, gao2015ins, soloviev2008tight, udacity_av_apollo, udacity_av_nd, coursera_av}.
%\vspace{-0.18cm}
\nsubsection{Lateral-Direction Localization Attack} \label{sec:background_msf_attack}
%\vspace{-0.1cm}
For AD localization, a direct threat is the attacks targeting the localization sensors such as GPS spoofing~\cite{zeng2018all, narain2018security, spoof_tesla, utaustin_russia_report, utaustinspoofer, kerns2014unmanned, popperccs11}, in which the attacker transmits fake satellite signals to the victim GPS receiver and thus cause it to resolve positions manipulated by the attacker. However, due to the high robustness provided by sensor fusion, MSF is often considered as a promising defense strategy for GPS spoofing~\cite{davidson2016controlling, lee2017attack, zeng2018all, cardenas2019cyber}. Contrary to the common belief, prior work~\cite{fusionripper} proposes an \textit{opportunistic} lateral-direction localization attack method, called \fr{}, which can use GPS spoofing alone to inject \textit{lateral deviations} in the MSF localization outputs and thus cause the AD vehicle to drive off-road or onto the wrong way. \fr{} is consist of two attack stages: \textit{vulnerability profiling} and \textit{aggressive spoofing}. In the vulnerability profiling stage, it spoofs the GPS inputs of MSF localization with a small constant distance $d$ (e.g., 0.5 m) in the lateral direction, waiting to discover a vulnerable attack window. Whenever the AD vehicle's physical deviation is larger than certain threshold (e.g., 0.3 m), \fr{} launches the aggressive spoofing stage, where a scaling factor $f$ (e.g., 1.2) will be continuously applied to the spoofing distance in each second to quickly introduce large lateral deviations in the MSF localization outputs. \fr{} has shown high attack effectiveness on the representative MSF algorithms, including the one in the industry-grade Baidu Apollo AD system~\cite{apollo}. To best of our knowledge, \fr{}~\cite{fusionripper} is the only localization attack that is able to defeat the MSF based localization algorithm in high-level AD systems.

\nsubsection{Threat Model} \label{sec:background_threat_model}

\textbf{Attacker's capability.}
In this work, we assume the attacker can launch practical lateral-direction localization attacks through external means such as GPS spoofing, which can cause lateral deviations in the localization outputs. Specifically, we focus on the lateral-direction attacks since such attacks (1) can cause the AD vehicle to violate the traffic norm that a vehicle should be driving within its designated lane boundaries and should not have unexpected lane straddling behaviors, and (2) pose a direct threat to the AD vehicle and road safety, e.g., it can cause the AD vehicle to drive off highway cliff or onto the wrong way and being hit by other vehicles that failed to yield in time. 

In particular, we do not consider simultaneous attacks that target both AD localization and lane detection at the same time, since such simultaneous attack neither already exists, nor can be easily achieved today (detailed discussions in \S\ref{sec:discuss}).
% due to the difficulty of attack synchronization and the non-deterministic nature of existing lateral-direction localization attack~\cite{fusionripper}. 

% We consider as out-of-scope for the attacks that can manipulate both the AD localization and perception outputs.

\textbf{AD control assumption.}
Same as \fr{}~\cite{fusionripper} and also as a common design in academia~\cite{paden2016survey} and industry~\cite{apollo, autoware}, we assume the AD systems are designed to drive at the center of traffic lane and constantly correct the deviations to the center. Since AD controllers constantly correct such deviations at a high frequency, e.g., 100 Hz~\cite{apollo}, the lateral deviations in the AD localization will thus be directly reflected as physical world deviations, but to the opposite direction.
% \vspace{-0.05in}
\nsection{Lane Detection for High-Level AD Localization Defense: Opportunity Analysis} \label{sec:opportunity}

\textbf{Motivation and novelty.} Currently, no software-based defense solutions have been proposed to address the latest GPS spoofing-based lateral-direction localization attack in high-level AD systems (\S\ref{sec:background_msf_attack}). The closest ones are the recent physical-invariants based detectors proposed for small robotic vehicles such as drones and rovers, e.g., SAVIOR~\cite{savior} and CI~\cite{ci}, which estimate the physical dynamics of drones and rovers to validate the GPS signal. 
Although they show high effectiveness for such small robotic vehicles under large attack deviation goals, their effectiveness in AD vehicle context is fundamentally more limited since (1) existing vehicle dynamics models have difficulties in modelling high-speed and curvy-road settings~\cite{kong2015kinematic, polack2017kinematic}; and (2) in the AD context, the attack deviation goals can be much smaller (thus harder to detect) while still being safety-critical. As we concretely evaluate later in \S\ref{sec:eval_detection}, direct adaptation of such existing physical-invariant based approach to the AD context suffers from very high false positives and is actually close to random guessing.

In comparison to small robotic vehicles, the AD context may also have its unique defense opportunities for such lateral-direction localization attacks. \textit{Lane Detection (LD)}~\cite{hillel2014recent, pan2018spatial}, a technology commonly used in low-level AD systems for lane centering~\cite{openpilot, autopilot}, is such an example that can be used to measure the vehicle's lateral position within the current lane in real time, which is directly related to the lateral-direction attack goal (lane departure). Although effective in low-level AD systems (e.g., Level-2 ones such as Tesla Autopilot~\cite{autopilot} that still count on human drivers to take over anytime), \textit{LD is currently not used for high-level AD localization purpose (e.g., Level-4 ones such as Waymo that do not assume onboard human drivers)}. This is because what LD can provide is by nature only \textit{local} positioning (i.e., relative positioning within ego lane), while high-level AD requires \textit{global} positioning (i.e., in world coordinates on a map) for safe and correct driving decision-making without human drivers. Although there exist camera-based global localization methods using lane markings~\cite{kang2020lane, evlampev2020map}, they are not generally adopted in state-of-the-art high-level AD localization~\cite{wan2018robust, gao2015ins, soloviev2008tight, udacity_av_apollo, udacity_av_nd, coursera_av} as they are far from reaching the required centimeter-level accuracy~\cite{levinson2007map, reid2019localization, ega_requirement_report}.
%not only require extra efforts for lane-map generation~\cite{}, but also 
% for high-level AD~\cite{levinson2007map, reid2019localization, ega_requirement_report}.
% To the best of our knowledge, \textit{no prior works have used LD to defend against localization attacks;
% and state-of-the-art high-level AD systems do not use LD for localization~\cite{wan2018robust, autoware, gao2015ins, soloviev2008tight, udacity_av_apollo, udacity_av_nd, coursera_av}. 

%nd (2) high-level AD localization requires centimeter-level localization accuracy~\cite{levinson2007map, reid2019localization, ega_requirement_report} for safe and correct driving without human drivers, while LD can only provide less-accurate lateral positioning~\cite{} and is incapable of longitudinal localization. 

While less suitable for global localization accuracy purposes in high-level AD,
in this paper we propose to be the first to explore novel use of LD for \textit{defense purposes} in high-level AD localization. To concretely understand the potential of such a domain-specific defense opportunity, we analyze LD's defense properties in the following 5 general aspects.

%\alfred{include argument on LD is not used for high-level Ad localization and why? I remember it is a common reviewer question.} \junjie{added.}

% Although seemingly promising, LD also has limitations that may hinder its defense practicality. 

\newparts{

\textbf{1) General to lateral-direction localization attack.}
As mentioned above, LD can provide real-time information directly related to the \textit{attack goal} of lateral-direction localization attacks. Thus, LD by nature has the potential to provide general defense capabilities to not only the existing attack designs such as those in~\S\ref{sec:background_msf_attack}, but also their potential adaptive versions or other new attack designs in the future, as long as the attack goal is to cause lateral deviations. 

%As mentioned above, LD is directly related to the attack goal of lateral-direction localization attacks. Therefore, LD-based defenses are general to any GPS spoofing methods that aim to cause lane departure, among which \fr{}~\cite{fusionripper} is the state-of-the-art and is so far the only effective one for MSF-based localization in high-level AD systems.
%\alfred{talk about the generality over fusionripper} \junjie{added.}

\textbf{2) Technology maturity.}
Benefit from the growing prosperity of Deep Neural Networks (DNNs), LD is already a mature technology that has been used for lane centering in low-level AD systems and vehicles, e.g., OpenPilot~\cite{openpilot}, Tesla Autopilot~\cite{autopilot}, GM Cadillac, Honda
Accord, Toyota RAV4, Volvo XC90, etc.
In fact, the existing camera-based LD solutions are quite robust to the dynamic environmental conditions. For example, Tesla Autopilot can effectively recognize lane lines even during a night storm~\cite{autopilot_night_rain}. 
Apart from DNN advancement, the camera auto-exposure and vehicle headlights also improve the usability of LD. Later in \S\ref{sec:eval}, we also evaluate our defense on datasets with various environmental conditions and show that it is robust to low visibility conditions.
}

\newparts{
\textbf{3) Defense deployability.}
Since today's high-level AD vehicles are all equipped with cameras for road object detection, using them for an LD-based defense solution is thus readily deployable without the need to install any new hardware. Moreover, many state-of-the-art LD models are publicly available~\cite{pan2018spatial, neven2018towards}, including those used in industry-grade lane centering systems~\cite{openpilot}; some high-level AD systems are also using LD for camera calibrations~\cite{apollo}.}
% Therefore, deploying LD to high-level AD systems typically will not impose technical challenges.

% Information is generally available in AD context

\newparts{
\textbf{4) Defense coverage.}
For LD to be effective, lane line markings are required, which may not be available in local road segments such as intersections. Interestingly, due to real-world sensor noises and algorithm inaccuracies, the attacks to MSF localization are \textit{fundamentally opportunistic}. For example, despite having a high overall attack success rate, latest lateral-direction localization attack cannot predict when and where a large deviation can be injected to the MSF outputs~\cite{fusionripper}.
Due to such opportunistic property, the attacker \textit{cannot deterministically cause a desired lateral deviation to only appear in regions without lane line markings}. Such an attack property is fundamental to the MSF localization designs popularly used in high-level AD systems, since with this design the attack effectiveness is fundamentally dependent on sensor noises and algorithm inaccuracies of other sources, which are neither observable nor controllable by a tailgating attacker~\cite{fusionripper}.

Motivated by this insight, we analyze all attack traces evaluated in the \fr{} paper~\cite{fusionripper} and our own evaluation later (\S\ref{sec:eval_method}), and find that LD can indeed provide a decent practical defense coverage: among all attack starting points in the traces, only \textit{0.8\% (15/1813)} achieved the attack goal in road regions without lane line markings. Thus, an LD-based defense, if effective, can already provide protection for the 99.2\% of the possible attack attempts. In addition, autonomous trucks, which are an important high-level AD application, are generally not subject to such limitation since they mainly operate on the ``middle mile'' (i.e., highways)~\cite{tusimple_ad_truck_middle_mile, walmart_ad_truck_middle_mile}, where lane line markings are generally always available. 
}

% FusionRipper attack success rate in areas without lane line markings: 
% 0.83\% = 15 / 1813 (among all attack traces)
% 1.41\% = 15 / 1062 (among all local attack traces)

\newparts{
\textbf{5) Independence to existing localization attack.}
To defend against existing attacks, a desired defense property is that the lane line markings perceived by LD are not already used in MSF localization. This is because if such information is already used, existing attacks might have already exploited their vulnerable periods (e.g., natural detection inaccuracies), making the additional use of such information for defense less likely to be effective. 
%make the defense robust and therefore practically usable.
In representative MSF localization designs, LiDAR locator is the only one among MSF inputs (\S\ref{sec:background_msf_attack}) that is possible to utilize lane line markings as features. Thus, we perform an experimental analysis to understand the dependency between state-of-the-art LiDAR locators~\cite{wan2018robust, autoware} and lane line markings in Appendix~\ref{app:lidar_lane_line_dependency}. Our results show that today's LiDAR localization algorithms have a \textit{statistically-strong independence} of the lane line markings, very likely because lane markings is much less useful for global localization on a map compared to more unique road features such as buildings, roadside layouts, and traffic signs. This thus suggests that LD can indeed provide independent defense information to existing attacks. However, such independence property will disappear in adaptive attack settings (i.e., consider attacking LD after the defense is deployed). Thus, we require our defense design to be fully-aware of such adaptive attack surface (\S\ref{sec:design_overview}), and also evaluate it later (\S\ref{sec:adaptive_attacks}).

%Admittedly, even with perfect independence between LD and MSF, attacks that target both can potentially bypass the defense. However, as will be discussed in \S\ref{sec:discuss}, such a simultaneous attack neither already exists, nor can be easily achieved.
% due to the difficulty of attack synchronization and non-determinism of the existing lateral-direction localization attack~\cite{fusionripper}.
%\alfred{is there a separate more detailed discussion on this? If so please cross-ref (again, no repetitive logic). In such detailed discussion, we should emphasize how fundamental such a property is. } \junjie{now refering to the limitation discussion section on simultaneous attack to MSF and LD.}
}

% After doing this, a new challenge is if an attacker is aware of this source, whether they can bypass it? We have evaluation referring to stealthy attack evaluation

% one adaptive attack is FusionRipper and LD attack together
% but fusion ripper is non-determinism, hard to do it together

% \vspace{-0.05in}
\nsection{Novel LD-based Defense Design: \ld{}} \label{sec:design}

% To leverage the defense-suitable information source and address challenges \textbf{C2} and \textbf{C3}, we propose 

Considering the multi-dimensional defense opportunities above, in this paper we are motivated to design the first domain-specific lane detection-based defense approach against lateral-direction AD localization attack, named \textit{\textbf{\ldi{}} (\underline{L}ane \underline{D}etection based \underline{L}ateral-\underline{D}irection \underline{L}ocalization attack \underline{D}efense)}. In this section, we first describe the associated design challenges and then present the design details.

%In this section, we present our defense design, \textbf{\ld{}}, which stands for \underline{L}ane \underline{D}etection based \underline{L}ateral-\underline{D}irection \underline{L}ocalization attack \underline{D}efense.
% Specifically, our design leverages the LD to perform both \textit{detection} and \textit{response} to the lateral-direction localization attacks targeting high-level AD systems.

\nsubsection{Design Challenges} \label{sec:design_challenges}

Although LD comes with various defense opportunities, systematically leveraging it for AD localization defense purposes still needs to address the following design challenges:

%can readily provide information directly related to lateral-direction attacks (i.e., lateral deviation to lane departure)

\textbf{C1: Non-trivial design details for attack detection.} Although at the high level LD can provide information directly related to lateral-direction attacks (i.e., lateral deviation to lane departure), at the detailed defense design level there are still many technical challenges we need to address, for example (1) incompatibility of the coordinate systems, i.e., LD is by default in \textit{local} positioning coordinate system (i.e., within the ego lane), while the attack is in \textit{global} coordinate system (i.e., the world coordinates); (2) choice of the attack-influenced information level for attack detection, e.g., directly at the spoofed GPS signal level or at the attack-influenced MSF output level; and (3) sufficient robustness to natural LD inaccuracies in practice, e.g., missing or incorrect detection, for minimizing possible false positives in attack detection.

%Since LD measures AD vehicle's relative position in the ego lane, at the high level it seems to be straightforward to use it to detect any unexpected lateral deviations in localization. However, we find there are still quite some non-trivial detailed design questions that are critical to the detection performance in practice, for example at which attack-influenced information level should the detection be performed, e.g., at the GPS signal or the MSF output level; and how to improve the robustness to natural LD inaccuracies, e.g., missing or incorrect detection.

%how to handle the incompatibility of the coordinate systems between LD and AD localization; 

%since LD is not readily used in high-level AD localization (\S\ref{sec:opportunity}), there are still many critical design  that can directly impact the detection effectiveness in practice, for example (1) incompatibility of the coordinate systems, for which LD is by default in \textit{local} positioning coordinates (i.e., within the ego lane), while the direct target of existing attacks is in \textit{global} coordinates (i.e., the world coordinates); (2) choices of the attack-influenced information level for attack detection, e.g., directly at the spoofed GPS signal level or the attack-influenced MSF output level; and (3) robustness to natural LD inaccuracies, e.g., missing or incorrect detection.

\textbf{C2: Need for AD-specific attack response design.} Since high-level AD vehicles are travelling at high speed and by design cannot assume on-board human driver ready for take-over at any time (already the case in some commercial AD services~\cite{waymo_driverless, baidu_driverless_robotaxi}), it is necessary to further design an attack response step that can (1) minimize the safety risks during response, and (2) assume no dependence on human assistance. For small robotic vehicles such as drones and rovers, prior works have considered using state estimation models to replace the attacked physical sensor after attack detection~\cite{choi2020software, zhang2020real}. However, such methods still count on human operators to take over as soon as possible since such state estimations cannot replace physical sensors for a prolonged duration due to drifting~\cite{choi2020software}, not to mention that such models are suffering from much more severe motion model accuracy limitations when applied to the AD context (\S\ref{sec:eval_detection}). Thus, a new design is needed to achieve our AD-specific response goal above.

%For small robotic vehicles such as drones and rovers, various prior works propose attack detection-only defense solutions, which are indeed already practically useful since the attack can be directly handled by emergency stop and/or human operators. However, in high-level AD settings, we find attack detection alone is not enough in practice, since AD vehicles are travelling in much higher speed and by default there does not exist on-board human driver ready for taking over the driving all the time. Thus, in AD context it is \textit{necessary} to design an automatic and safe attack response method following up the attack detection.

\textbf{C3: Adaptive attack from LD side.} While LD is currently independent to existing high-level AD localization attacks due to the lack of use (\S\ref{sec:opportunity}), our defense-purpose use of it in \ld{} is inherently introducing a new attack surface from the LD side. In fact, recent works have already discovered concrete lateral-direction attacks against LD in production AD context~\cite{sato2021dirty}. To systematically account for such inherent adaptive attack surface, our defense design thus needs to consider the more challenging setup where both the attack detection and response designs cannot simply assume the LD side is trustworthy (and use it as the benign reference accordingly) when its outputs are inconsistent with the AD localization side.

%; instead, we need to consider the more challenging setup that both sides 

\cut{

thus needs to further avoid simply taking the LD-side inputs as the benign value when they are in conflict

faces the challenge that we cannot simply trust the LD-side input in both attack detection and response 

when it is in conflict with the localization

The usage of LD in the defense unavoidably exposes it to potential LD-side attacks, e.g., DRP attack~\cite{sato2021dirty}. Therefore, it is necessary to design the defense such that it is also resilient to the new threats.

\ld{} leverages the camera-based LD to facilitate both attack \textit{detection} and \textit{response}. 
Since LD measures AD vehicle's relative position in the traffic lane, it is thus straightforward to use it to detect any unexpected lateral deviations in localization.
However, to utilize LD for defense purposes, we recognize several design \textit{challenges}, for which we need to weigh between potential solutions to determine a practical defense design.
\vspace{-\topsep}
\begin{itemize}
\setlength{\itemsep}{0pt}
\setlength{\parskip}{0pt}
    \item \textit{C1: At which attack-influenced information level should we perform the detection?} Since existing lateral-direction localization attack leverages GPS spoofing as an attack vector to affect the MSF outputs, one has to determine whether to perform the detection at the GPS or at the MSF output level.
    \item \textit{C2: Besides attack detection, an AD-specific attack response design is required.} Although there exists prior attack recovery methods~\cite{choi2020software, zhang2020real} for drones and rovers, which use state estimation models as sensor replacement during recovery, their methods suffer from the same model accuracy issue as in SAVIOR~\cite{savior} and CI~\cite{ci} (more details in \S\ref{sec:eval_detection}). In addition, they often intend to maintain a normal system operation during attack recovery and assume a human driver would take over the control afterwards. However, this is not the case for high-level AD vehicles, where typically no safety drivers will be onboard once commercially deployed~\cite{baidu_driverless_robotaxi, waymo_driverless}. Lastly, comparing to drones and rovers, the more complex driving environment and higher driving speeds demand an AD-specific response design that reflects the safety-first principle~\cite{ad_safety_first}. 
    % Thus, a systematic attack response design is especially necessary for high-level AD systems.
    \item \textit{C3: How to make the defense resilient to adaptive attacks that target LD?} The usage of LD in the defense unavoidably exposes it to potential LD-side attacks, e.g., DRP attack~\cite{sato2021dirty}. Therefore, it is necessary to design the defense such that it is also resilient to the new threats.
    % Therefore, it is important to design the defense
    \vspace{-\topsep}
\end{itemize}
}

\vspace{-0.05in}
\nsubsection{Design Overview} \label{sec:design_overview}
% \vspace{0.05in}

In this section, we explain each design component in \ld{} and how they address above design challenges. Fig.~\ref{fig:design_overview} shows an overview of \ld{} fitted in a typical high-level AD system.

\begin{table*}[tbp]
\footnotesize
\begin{minipage}{0.58\linewidth}
	\centering
    \includegraphics[width=\columnwidth]{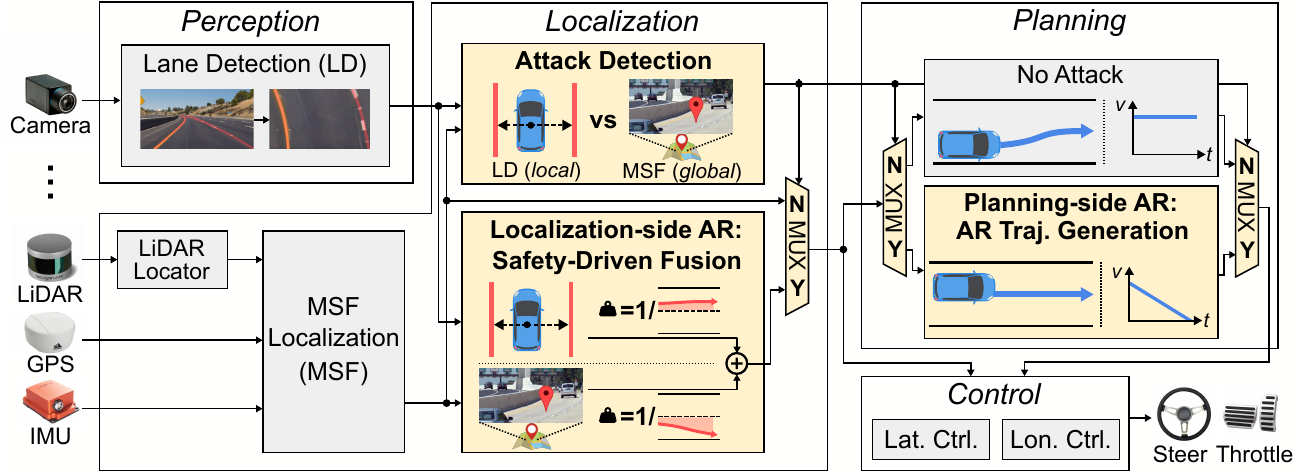}
    % \vspace{-0.25in}
    \captionof{figure}{Overview of \ld{} design integrated in a typical high-level AD system. New components are highlighted in yellow.}
    \label{fig:design_overview}
\end{minipage}\hfill
\begin{minipage}{0.39\linewidth}
	\centering
    \includegraphics[width=\columnwidth]{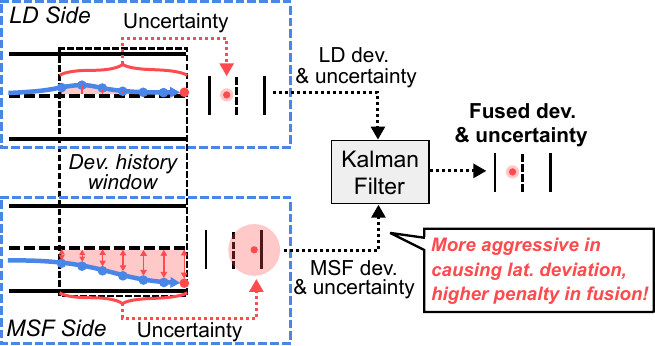}
    % \vspace{-0.25in}
    \captionof{figure}{Illustration of safety-driven fusion in the attack response (\S\ref{sec:design_ar}).}
    \label{fig:safety_driven_fusion}
\end{minipage}
\vspace{-0.2in}
\end{table*}

\cut{
\begin{figure}[tbp]
\centering
\includegraphics[width=\linewidth]{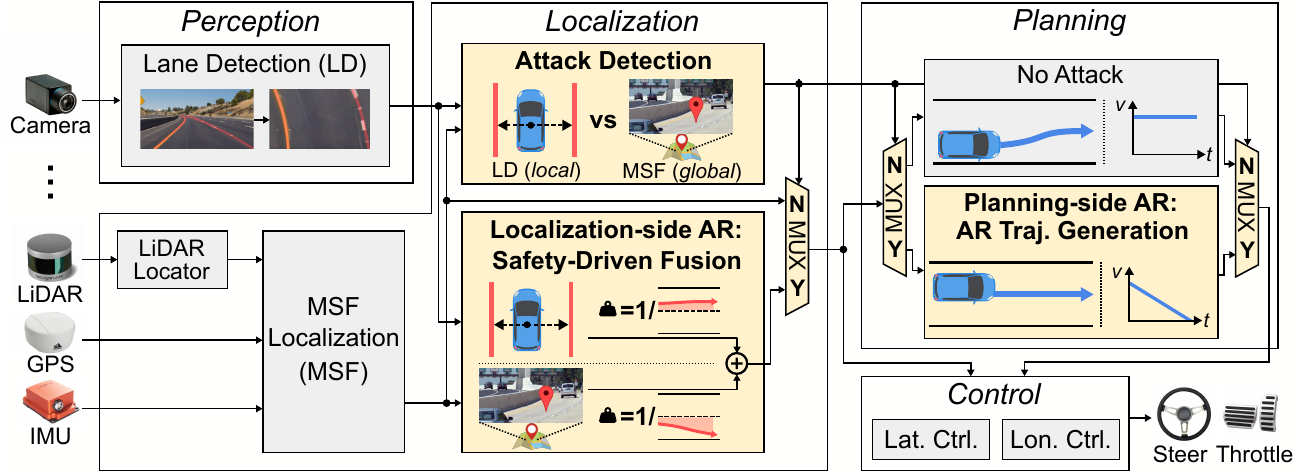}
\vspace{-0.25in}
\caption{Overview of \ld{} design integrated in a high-level AD system. New components are highlighted in yellow.}
\label{fig:design_overview}
\vspace{-0.05in}
\end{figure}
}

% \textbf{Attack detection via MSF/LD cross-checking.}
\textbf{Attack detection at MSF output level.} As shown in Fig.~\ref{fig:design_overview}, the attack detection step is performed in the localization module to constantly check the consistency between the LD outputs and original localization output and raise anomalies use popular anomaly detectors such as CUmulative SUM (CUSUM). To address the incompatibility of their coordinate systems mentioned in \textit{C1}, we convert both into a unified lateral deviation representation w.r.t. the \textit{lane centerline} since that's directly related to the lateral-direction attack goal. Regarding the choice of the attack-influenced information level for attack detection, we choose to detect at the MSF output level rather than at the GPS output level since (1) in normal conditions, GPS positions can naturally have large noises while MSF outputs are at centimeter-level accuracy~\cite{wan2018robust}. Thus, performing the detection at the MSF level can better reduce false positives; and (2) detecting at the MSF output level also allows \ld{} taking advantage of the \textit{opportunistic} property of \fr{}, for which the attacker cannot predict where and when MSF will exhibit large deviations. This thus can make it much more difficult for the attacker to easily bypass the detection by targeting locations without lane line markings. We also have designs for addressing false positives from common lane detection inaccuracies.

\textbf{Attack response via safe in-lane stopping.}
As discussed in \textit{C2}, we need a new AD-specific design for the Attack Response (AR) step. There are several common choices in human driving if the vehicle navigation is malfunctioning, for example maintaining driving in the current lane waiting for the system to recover, or pulling over to the road side. However, these cannot apply to the context of AD localization attacks, since without knowing the accurate real-time location, we cannot even know how to safely and correctly drive in the current lane or to road side. We also cannot blindly count on the LD outputs to drive due to the need to account for the adaptive attack surface on the LD side (\textit{C3}). Thus, we consider the safest AR choice is to try to safely stop in the current lane, which has the minimal reliance on the attack-time localization accuracy for maximizing safety in the AR period. More importantly, on the attack side, since this minimizes the attackable duration after detection, it can fundamentally \textit{bound} the attack-achievable deviation in AR period. Even though stopping in the ego lane is not ideal, it is commonly recognized~\cite{nhtsa_standard} as one of the fallback strategies to transition to a minimal risk condition when the AD vehicle cannot operate safely. In most driving scenarios, stopping in the ego lane shall not cause a collision as long as the tailgating vehicle is driving with safe following distance and speed, which is much safe than driving out of the ego lane.

\textbf{Safety-driven fusion for adaptive LD attack.} Although the in-lane stopping AR strategy can already bound the attack-achievable deviation, it is still highly desired if we can minimize the attacker's impact on the localization accuracy during the AR period, since to safely stop, there is still a long stopping distance that the ego vehicle has to travel, especially when the speed is high (e.g., over 50 meters at 60 mph~\cite{stopping_distance}). To account for the adaptive LD attack surface (\textit{C3}), the key challenge is how to decide which side (LD or MSF) to trust when they are conflicting with each other in AR. Motivated by the safety-first principle in production AD design~\cite{ad_safety_first}, we propose a novel \textit{safety-driven fusion} design, which systematically decides the contributions from different fusion inputs based on their tendencies to cause unsafe driving; the higher such tendency is, the smaller their contributions will be to the final fusion output. In our problem context, such a tendency is judged by the deviation aggressiveness to cause lane departure, which will thus by default penalize the attacked side no matter it is from LD or MSF, leading to less attack-introduced deviation. To bypass this penalty and more effectively influence the fused results, the attacked side has to be less aggressive in lateral deviations. However, given the limit on the attackable duration imposed by the in-lane stopping AR strategy, the attack-achievable deviation during AR will still be reduced. Thus, under our AR design that bounds the attackable duration, such safety-driven fusion design can further fundamentally reduce the attacker's capability in causing safety damages during AR even in adaptive settings.

\vspace{-0.05in}
\nsubsection{Attack Detection Design} \label{sec:design_detection}
% \vspace{0.05in}

\begin{algorithm}[tbp]
\footnotesize
\caption{Attack detection by checking the consistency between MSF and LD}
\label{alg:attack_detection}
\textbf{Notations:} $MSF$: MSF position output; $LD$: lane detection output; $S$, $b$, $\tau$: CUSUM statistic, weight, anomaly threshold; $D$: deviation to lane centerline; $lw_{\text{map}}$: lane width from semantic map\\
\textbf{Initialize:} $S_0\gets 0$
\begin{algorithmic}[1]
\ForEach{new lane detection output $LD_i$} \Comment{e.g., runs at 20Hz} 
    \State $MSF_i \gets \text{latest MSF position}$ \Comment{MSF is often more frequent than LD}
    \State $D_i^{\text{MSF}}$ $\gets$ \text{M\textsc{ap}}\text{L\textsc{ane}}\text{D\textsc{ev}}($MSF_i$) \Comment{MSF dev. (Appendix~\ref{app:map_apis})}
    \State $lw_{\text{map}}$ $\gets$ \text{M\textsc{ap}}\text{L\textsc{ane}}\text{W\textsc{idth}}($MSF_i$) \Comment{lane width (Appendix~\ref{app:map_apis})}
    \State $D_i^{\text{LD}}$ $\gets$ \text{L\textsc{d}}\text{D\textsc{ev}}($LD_i$, $lw_{\text{map}}$) \Comment{LD dev. to centerline (Alg.~\ref{alg:ld_dev_calc})}
    \State $S_i \gets$ max(0, $S_{i-1} + \lvert D_i^{\text{MSF}} - D_i^{\text{LD}} \rvert - b$) \Comment{calc. CUSUM statistic}
    \If{$S_i > \tau$}
        \State \text{attacked} $\gets \text{true}$ \Comment{report under attack if over threshold}
        \State \textbf{break}
    \EndIf
\EndFor
\State $\Rightarrow$ \textit{switch to attack response}
\end{algorithmic}
\end{algorithm}

As described above, we choose to perform the attack detection at the MSF output level, which is thus designed as a post-processing step in the localization module as shown in Fig.~\ref{fig:design_overview}.
%we leverage the lateral deviations perceived by LD to perform the attack detection, which
% the attack detection steps will introduce two new components to the high-level AD systems: \textit{lane detection (LD)} and \textit{attack detection}.
% \textbf{Attack detection.}
%the attack detection is designed as a post-processing step in the localization module to constantly check the consistency between the MSF and LD outputs. 
% Since MSF and LD usually operates at different frequencies, e.g., Baidu Apollo runs MSF localization at 100 Hz and LD at 20 Hz, we select the less frequent one to trigger the attack detection.
The detection algorithm is shown in 
Alg.~\ref{alg:attack_detection}.
% Since MSF outputs positions in the global coordinate system and LD measures the relative distance of the vehicle to the lane lines in the local frame, their results are not directly comparable. 
As mentioned in \textit{C1}, the MSF and LD outputs are in different coordinate systems. 
Therefore, we first need to convert them to a unified coordinate system such that they are comparable. 
% the MSF outputs to the same local frame as the LD using the semantic map in high-level AD systems. 
For MSF outputs, we obtain an \textit{MSF-based lateral deviation to the lane centerline} ($D_i^{MSF}$ in Alg.~\ref{alg:attack_detection}) by querying the MSF position in the semantic map~\cite{lyft_semantic_map}, which is a standard utility on high-level AD systems storing the road geometry information of the area that the AD vehicle is allowed to drive.
% Appendix~\ref{app:map_apis} lists the semantic map APIs required in \ld{} that is generally available in high-level AD systems. 
For the LD outputs, we can calculate the lateral deviation to the centerline based on the left and right lane line polynomial functions (detailed in Appendix~\ref{app:design_impl}). 
However, real-world lane markings can be complicated and confusion sometimes. For example, it is common to find that one of the lane lines missing or incorrectly detected in regions with lane splitting and merging. Therefore, we design two optimizations to calculate a more robust lateral deviation from the LD outputs leveraging the lane width from the semantic map (detailed in Appendix~\ref{app:design_impl}), which is a problem-specific improvement opportunity since in the main LD usage domain, low-level AD systems, such semantic maps are not generally available.
Since \ld{} relies on the existence of lane line markings, we disable the attack detection prior to entering these regions based on the information from the semantic map.

% Specifically, to handle cases when one of the lane lines is missing or incorrectly detected, we apply two optimizations to calculate a more robust lateral deviation implementation details in \S\ref{sec:design_impl}.

% After obtaining the MSF- and LD-based lateral deviations, we can then use their deviation consistency to determine if the MSF localization is under attack. This is because the attack-introduced lateral deviation in the MSF results in a physical world deviation to the opposite direction, which will be observed by the LD. While in the benign cases, the lateral deviation in the MSF is a natural outcome of the vehicle's physical world lateral movement, and thus the lateral deviations from LD and MSF will be consistent.

After obtaining the MSF- and LD-based lateral deviations, we can then use their deviation consistency to determine if MSF localization is under attack. To do so,
% To check the consistency between the MSF and LD lateral deviations, 
we apply the widely-used CUSUM anomaly detector (line 6--10 in Alg.~\ref{alg:attack_detection}),
which has shown high detection effectiveness in prior works~\cite{urbina2016limiting, savior}.
The CUSUM detector calculates a statistic $S_{i} = max(0, S_{i-1} + \lvert r_i \rvert - b); S_0 = 0$, where $r_i = D_i^{\text{MSF}} - D_i^{\text{LD}}$ is the residual between the MSF and LD lateral deviations, $b$ is a weight to prevent the CUSUM statistic from monotonically increasing in the benign scenarios. We consider as under attack if $S_i$ is over a certain threshold $\tau$.
Once an attack is detected, we then switch to the Attack Response stage.

\vspace{-0.05in}
\nsubsection{Attack Response Design} \label{sec:design_ar}
% \vspace{0.05in}

% The Attack Response (AR) stage is separated into two components, with one in the localization module and another in the planning module as shown in Fig.~\ref{fig:design_overview}. In the current AR design, we focus on achieving the most basic AR goal to stop within the current lane boundaries.

As described in~\S\ref{fig:design_overview}, we consider safe in-lane stopping as the safest AR choice. 
%Besides basic emergency operations such as turning on hazard warning lights to signal surrounding vehicles, the AR stage is designed to properly handle our attack response goal: safely stop in the current lane (\S\ref{sec:design_overview}). 
As shown in Fig.~\ref{fig:design_overview}, the AR is composed of two components to safely drive the vehicle before stop: (1) AR trajectory generation on the planning side, and (2) safety-driven fusion of MSF and LD on localization side.

% also need a systematic AR design to control the vehicle to stay in the current lane. In high-level AD systems, the longitudinal and lateral controllers calculate throttling and steering commands based on the current position and a planned trajectory. However, during AR, we can no longer purely rely on the MSF localization for positioning and the planned trajectory also needs to be updated to reflect the intended AR trajectory. Thus, we design a \textit{safety-driven fusion} component in the localization module and an \textit{emergency trajectory generation} component in the planning module as shown in Fig.~\ref{fig:design_overview} to achieve our AR goal: stopping within the current lane boundaries.

\textbf{Planning-side AR: AR trajectory generation.} 
%To realize the AR strategy, we need to design an AR planning trajectory such that the lateral and longitudinal controllers can use as reference to enforce the AR goal.
%Since our AR goal is to stop in the ego lane, we design the AR trajectory to be aligned with the lane centerline. To reduce the speed, we then set a slowing-down speed profile on the AR trajectory based on a safe deceleration value used in high-level AD systems.
The planning module in the high-level AD system periodically generates planned trajectories, which the controllers take as speed and lateral position references to produce throttling and steering commands. Thus, to enforce the AR goal, the planning module needs to generate an AR trajectory with a stopping motion. 
Since our AR goal is to stop in the ego lane, we designed the AR trajectory to be aligned with the lane centerline. To reduce the speed, we then set a slowing-down speed profile on the AR trajectory based on a safe deceleration value used in high-level AD systems. Generally, a deceleration $<$4.6 $\mathrm{m/s^2}$ is considered as safe for maintaining steady control~\cite{deceleration}. Thus, to calculate the speed profile of the AR trajectory, we apply 4 $\mathrm{m/s^2}$ as deceleration, which is also defined in Baidu Apollo as the maximum allowed deceleration to ensure safety~\cite{apollo}. 
%To do that, we generate a speed profile of the AR trajectory using the maximum safe deceleration for vehicles (Appendix~\ref{app:design_impl}). 
% Generally, a deceleration $<$4.6 $\mathrm{m/s^2}$ is considered as safe for maintaining steady control~\cite{deceleration}. Thus, we apply 4 $\mathrm{m/s^2}$ as the deceleration, which is also defined in Baidu Apollo as the maximum allowed deceleration to ensure safety~\cite{apollo}. 
Note that since the original planning algorithms are typically designed under the assumption that the localization accuracy is high (i.e., cm-level~\cite{reid2019localization}), we find directly re-using such algorithms in AR will result in unstable control since the planned trajectories are too sensitive to the larger localization errors and uncertainties after fusing the LD and MSF sides when one side is under attack. Thus, we directly set the planned trajectory as the centerline of the ego lane to achieve more stable control.

\begin{algorithm}[tbp]
\footnotesize
\caption{Safety-driven fusion for attack response}
\label{alg:ar_fusion}
\textbf{Notations:} $D$: deviation to lane centerline; $P$: uncertainty from MSF or LD outputs; $MSF$: MSF position output; $kf$: 1-dimensional Kalman Filter;  $R$: uncertainty for KF update
\begin{algorithmic}[1]
\Function{\text{F\textsc{used}}\text{P\textsc{ose}}}{$D^{\text{MSF}}$, $D^{\text{LD}}$, $P^{\text{MSF}}$, $P^{\text{LD}}$, $MSF$}
    \State $R^{\text{MSF}}, R^{\text{LD}} \gets \text{U\textsc{ncertainty}}(D^{\text{MSF}}$, $D^{\text{LD}}$, $P^{\text{MSF}}$, $P^{\text{LD}})$
    \State $kf.update(D^{\text{MSF}}, R^{\text{MSF}})$; $d \gets kf.predict()$
    \State $kf.update(D^{\text{LD}}, R^{\text{LD}})$; $d \gets kf.predict()$
    \State $pose_{\text{center}}$, $heading_{\text{center}}$ $\gets \text{M\textsc{ap}}\text{L\textsc{ane}}\text{P\textsc{oint}}(MSF)$ \Comment{Appendix~\ref{app:map_apis}}
    \State $pose_{\text{fusion}}$ $\gets \text{A\textsc{dd}}\text{D\textsc{ev}}\text{T\textsc{o}}\text{P\textsc{oint}}(pose_{\text{center}}$, $heading_{\text{center}}, d)$
    \State \textbf{return} $pose_{\text{fusion}}$
\EndFunction
\end{algorithmic}
\end{algorithm}

\textbf{Localization-side AR: safety-driven fusion.}
% Since the attack can happen on either MSF or LD side, we cannot simply fall back to an LD-based Automated Lane Centering lateral control design as in lower-level AD systems~\cite{openpilot}. Instead, we address this by applying a 
% As mentioned in \S\ref{sec:design_overview}, we design a \textit{safety-driven fusion} on the MSF/LD outputs to \textit{leverage the safer source} and \textit{penalize the more aggressive one} in the driving context as shown in Alg.~\ref{alg:ar_fusion}. Specifically, 
As described in \S\ref{sec:design_overview}, we need to design a safety-driven fusion algorithm on the localization side that can systematically fuse LD and MSF outputs while taking less contributions from the side that is more aggressive in causing lateral deviations. To achieve this, we leverage a classic fusion algorithm design, \textit{Kalman Filter} (KF) based fusion, which can systematically determine the contributions of each fusion source using \textit{uncertainties}~\cite{thrun2005probabilistic, friedland2012control}. In the original design, the uncertainty score calculation are based on the noise-level measurements reported by the sources themselves, which thus are not suitable in attack settings since such measurements are also fundamentally under the attacker's control.

To systematically realize our safety-driven fusion design between LD and MSF, we thus still leverage such uncertainties-based fusion framework but design novel uncertainty score calculation based on their tendencies to cause lane departure. 
%As briefly mentioned in \S\ref{sec:design_overview}, we use a 1-dimensional Kalman Filter (KF) to fuse the two sources (MSF and LD), and assign different uncertainty scores based on their safety implication in the driving context to indicate their trustworthiness.
% Due to the consistency-checking based detection design (\S\ref{sec:design_detection}), it is not straightforward to assign the uncertainties since we cannot tell which source is the malicious one.
% Since we cannot purely trust uncertainty output from the MSF/LD algorithms since they could be influenced by the attack, we thus design a novel uncertainty calculation, which uses the \textit{cumulative lateral deviations} on each source to penalize the more aggressive one during the fusion. 
% The intuition is that with such a design, the malicious source needs to be less aggressive in order to gain higher trust in the fusion. Yet, less aggressive means that the lateral deviation it can cause in the AR period will also be limited. 
Alg.~\ref{alg:ar_uncertainty} lists the pseudocode for the uncertainty calculation. As shown (lines 2 and 7), we store the historical lateral deviations from MSF and LD in two fixed-size windows. To obtain the uncertainties, we first calculate the cumulative deviations in these two windows, and then calculate their proportions to the geometric mean of them (lines 8 and 11). We choose geometric mean over arithmetic mean since it can better penalize the source with a larger cumulative deviation. To increase the design flexibility, we include both our cumulative lateral deviation based uncertainty and the uncertainty from MSF/LD algorithms in the final uncertainty and use a weight $\lambda$ to adjust their fractions (lines 12 and 13).

With the uncertainties, we apply standard KF update/predict operations to fuse the MSF and LD lateral deviations (lines 3 and 4 in Alg.~\ref{alg:ar_fusion}). 
% To create the global position required by the AD planning and control, 
We then add the fused lateral deviation to the closest centerline point along the lateral direction based on the lane heading to instantiate a fused localization in the global coordinate system (line 5--6 in Alg.~\ref{alg:ar_fusion}). Fig.~\ref{fig:safety_driven_fusion} illustrates an example of the safety-driven fusion process.

\begin{algorithm}[tbp]
\footnotesize
\caption{Cumulative lateral deviations based uncertainties calculation}
\label{alg:ar_uncertainty}
\textbf{Notations:} $D$: deviation to lane centerline; $P$: uncertainty from MSF or LD outputs; $DS$: deviation history; $w$: deviation history window size; $\lambda$: weight of the deviation history based uncertainty\\
\textbf{Initialize:} $DS^{\text{MSF}} \gets \{ \}$; $DS^{\text{LD}} \gets \{ \}$
\begin{algorithmic}[1]
\Function{\text{U\textsc{ncertainty}}}{$D^{\text{MSF}}$, $D^{\text{LD}}$, $P^{\text{MSF}}$, $P^{\text{LD}}$}
    \State $DS^{\text{MSF}}$ $\gets$ $DS^{\text{MSF}}$ $||$ $\lvert D^{\text{MSF}} \rvert$ \Comment{append MSF dev. to the history}
    \State $DS^{\text{LD}}$ $\gets$ $DS^{\text{LD}}$ $||$ $\lvert D^{\text{LD}} \rvert$ \Comment{append LD dev. to the history}
    \If{size of $DS^{\text{MSF}} > w$} \Comment{remove first element if full}
        \State $DS^{\text{MSF}} \gets DS^{\text{MSF}} \setminus DS^{\text{MSF}}[0]$
        \State $DS^{\text{LD}} \gets DS^{\text{LD}} \setminus DS^{\text{LD}}[0]$
    \EndIf
    \State $s^{\text{MSF}} \gets \sum_{n=1}^{w} DS^{\text{MSF}}$; $s^{\text{LD}} \gets \sum_{n=1}^{w} DS^{\text{LD}}$ \Comment{dev.  sums}
    \State $s^{GeoMean} \gets \sqrt{s^{\text{MSF}} \cdot s^{\text{LD}}}$ \Comment{geometric mean of dev. sums}
    \State $f^{\text{MSF}} \gets s^{\text{MSF}} / s^{GeoMean}$ \Comment{dev. history based uncertainty for MSF}
    \State $f^{\text{LD}} \gets s^{\text{LD}} / s^{GeoMean}$ \Comment{dev. history based uncertainty for LD}
    \State $R^{\text{MSF}} \gets \lambda f^{\text{MSF}} + (1 - \lambda) P^{\text{MSF}}$ \Comment{MSF uncertainty}
    \State $R^{\text{LD}} \gets \lambda f^{\text{LD}} + (1 - \lambda) P^{\text{LD}}$ \Comment{LD uncertainty}
    \State \textbf{return} $R^{\text{MSF}}, R^{\text{LD}}$
\EndFunction
\end{algorithmic}
\end{algorithm}

% \vspace{-0.05in}
\nsection{Defense Effectiveness Evaluation} \label{sec:eval}

In this section, we evaluate \ld{} against the state-of-the-art lateral-direction attack targeting high-level AD localization. 

\vspace{-0.05in}
\nsubsection{Evaluation Methodology} \label{sec:eval_method}
% \vspace{0.05in}

\textbf{Targeted AD system and attack.}
Since \ld{} is designed for high-level AD systems, we choose the industry-grade full-stack Baidu Apollo AD system~\cite{apollo} as a representative prototyping target. 
Specifically, Baidu Apollo adopts an MSF-based localization highly representative in both design (KF-based MSF) and implementation (state-of-the-art localization accuracy~\cite{wan2018robust}).
%a most representative MSF based localization design
% , which has been tested on a large AD vehicle fleet including various challenging driving scenarios such as urban roads, highways, tunnels, etc., 
%with state-of-the-art localization accuracy~\cite{wan2018robust}, 
% Baidu is also a leading AD developer currently already running commercial RoboTaxi services without safety drivers~\cite{baidu_driverless_beijing}.
Note that although our evaluation here uses Baidu Apollo, the \ld{} design itself is generalizable to other industry-grade high-level AD systems; for example, later in~\S\ref{sec:pixkit_eval} we also implemented it in Autoware for end-to-end physical evaluation.
% and will expand the business to 30 cities within next 3 years~\cite{baidu_driverless_robotaxi}.
For targeted attacks,
we evaluate against the recent \fr{} attack~\cite{fusionripper} since it is (1) the state-of-the-art and only lateral-direction localization attack that can break MSF localization; and (2) directly applicable to the above representative MSF implementation.

\textbf{Real-world sensor traces and \fr{} attack effectiveness.}
Since our evaluation target is the \fr{} attack, 
we follow the same evaluation methodology as in their paper~\cite{fusionripper} and conduct our evaluation on real-world sensor traces from the KAIST complex urban dataset~\cite{jeong2019complex}. Specifically, we look for traces with camera data as required by \ld{}, select the ones that the Apollo MSF can stably operate without attack~\cite{fusionripper}, and apply \fr{} from each consecutive timestamp as in~\cite{fusionripper}. In total, we obtain 562 attack traces summarized in Table~\ref{tbl:kaist_traces}. These traces cover diverse driving scenarios, e.g., different road types (344 on local roads and 218 on highways), driving speeds (9.5 to 26.3 m/s), time-of-day (e.g., 36 in the morning, 182 around sunset time), and road conditions (e.g., 170 with snow on road). 

For each trace, we follow the same method to identify the most effective attack parameters as in the \fr{} paper. Note that in the table our attack goal deviation is larger than the original \fr{} paper since they focus on the minimum urban lane width (i.e., 2.7 m) while we set the attack goal in a more realistic setting by measuring the lane widths in the dataset. This does not affect the attack effectiveness; as shown, the overall attack success rate is over 98\%, which is consistent with the \fr{} paper. In our evaluation, we exclude the scenarios without lane markings (e.g., when the vehicle is in an intersection) since it is out of the applicable domain for \ld{}. As analyzed in~\S\ref{sec:opportunity}, the lack of coverage of such scenarios do not eliminate the defense value since only 0.8\% of the attacks can possibly succeed in such scenarios and such successes are out of the attacker's control.

\begin{table}[tbp]
\centering
\footnotesize
\caption{Details of the 562 total attack traces used in our evaluation and the \fr{} attack effectiveness.}
% \vspace{-0.15in}
\label{tbl:kaist_traces}
\setlength{\tabcolsep}{2pt}
\begin{tabular}{@{}cccccccc@{}}
\toprule
\multirow{2}{*}{} & \multirow{2}{*}{\begin{tabular}[c]{@{}c@{}}Attack \\ Trace \#\end{tabular}} & \multirow{2}{*}{\begin{tabular}[c]{@{}c@{}}Road \\  Type\end{tabular}} &  \multirow{2}{*}{\begin{tabular}[c]{@{}c@{}}Avg. \\ Speed\end{tabular}} & \multicolumn{4}{c}{FusionRipper Attack} \\ \cmidrule(l){5-8} 
 &  &  &  & \begin{tabular}[c]{@{}c@{}}Attack\\ Goal Dev\end{tabular} & \begin{tabular}[c]{@{}c@{}}Best\\ $d$\end{tabular} & \begin{tabular}[c]{@{}c@{}}Best\\ $f$\end{tabular} & \begin{tabular}[c]{@{}c@{}}Success\\ Rate\end{tabular} \\ \midrule
\textit{ka-local31} & 174 & Local & 10.9 m/s & 1.3 m & 0.5 & 1.2 & 99.4\% \\
\textit{ka-local33} & 170 & Local & 9.5 m/s & 1.3 m & 0.3 & 1.3 & 98.3\% \\
\textit{ka-highway36} & 182 & Highway & 26.3 m/s &  1.9 m & 0.3 & 1.3 & 100\% \\
\textit{ka-highway18} & 36 & Highway & 24.8 m/s & 1.9 m & 0.3 & 1.3 & 100\% \\ \bottomrule
\end{tabular}
% \vspace{-0.05in}
\end{table}

\cut{
\begin{table}[tbp]
\centering
\footnotesize
\caption{Real-world sensor traces used in our evaluation and \fr{} attack effectiveness.}
\vspace{-0.1in}
\label{tbl:kaist_traces}
\setlength{\tabcolsep}{2pt}
\begin{tabular}{@{}cccccccc@{}}
\toprule
\multirow{2}{*}{Trace} & \multirow{2}{*}{\begin{tabular}[c]{@{}c@{}}Road\\ Type\end{tabular}} & \multirow{2}{*}{\begin{tabular}[c]{@{}c@{}}Avg.\\ Speed\end{tabular}} & \multirow{2}{*}{Duration} & \multicolumn{4}{c}{FusionRipper Attack} \\ \cmidrule(l){5-8} 
 &  &  &  & \begin{tabular}[c]{@{}c@{}}Attack\\ Goal Dev\end{tabular} & \begin{tabular}[c]{@{}c@{}}Best\\ $d$\end{tabular} & \begin{tabular}[c]{@{}c@{}}Best\\ $f$\end{tabular} & \begin{tabular}[c]{@{}c@{}}Success\\ Rate\end{tabular} \\ \midrule
\textit{ka-local31} & Local & 10.9m/s & 1014s & 1.3m & 0.5 & 1.2 & 99.4\% \\
\textit{ka-local33} & Local & 9.5m/s & 1283s & 1.3m & 0.3 & 1.3 & 98.3\% \\
\textit{ka-highway36} & Highway & 26.3m/s & 352s & 1.9m & 0.3 & 1.3 & 100\% \\
\textit{ka-highway18} & Highway & 24.8m/s & 162s & 1.9m & 0.3 & 1.3 & 100\% \\ \bottomrule
\end{tabular}
\end{table}
}

%\textbf{Semantic map creation.}
%Since the KAIST dataset 
% is designed for AD localization and perception tasks, it 
%does not provide any semantic maps,
% which are mainly used for AD planning~\cite{jeong2019complex}. As required in \ld{}, 
%we thus manually create the semantic map for each trace (more details in Appendix~\ref{app:map_creation}). \alfred{how important is this?} \junjie{
%although for KAIST traces, reviewer may not realize such a problem even if we delete this part, however, I think some of them may notice in the night-time driving trace evaluation. That's why I intended to keep this.}

\textbf{Lane detection and AD control effects under attack.} Since \ld{} does not assume any specific requirement on the lane detector, we are free to use any state-of-the-art lane detector or even an ensemble of lane detectors. In our evaluation, we opt to the LD model used in OpenPilot~\cite{openpilot}, which is already used commercially for Automated Lane Centering. The KAIST traces include time-synchronized left and right camera frames from a front-facing stereo camera. In our evaluation, we regard the left and right cameras as independent cameras and run the LD model and calculate lateral deviations (Alg.~\ref{alg:ld_dev_calc}) separately on them. We then aggregate their results to obtain an averaged lateral deviation on the LD side.

Since KAIST traces are collected under benign driving, we need to model the LD outputs when the AD localization is under attack. Same as the \fr{} paper~\cite{fusionripper}, we assume the lateral deviations in the MSF localization will be directly reflected as physical world deviations to the \textit{opposite} direction (\S\ref{sec:background_threat_model}). 
% since the AD control is actively correcting the MSF deviations at high frequencies (e.g., 100 Hz in Baidu Apollo~\cite{apollo}). 
We then model the attack-influenced LD outputs by adding the physical world deviations to the lateral deviations calculated from the benign LD outputs. Later in \S\ref{sec:end_to_end_eval}, we evaluate \ld{} in both end-to-end simulation and physical-world environments without such an assumption.

\textbf{Baseline: SAVIOR.} 
As a baseline, we evaluate the attack detection effectiveness of the closest alternative software-based method based on latest prior works for small robotics vehicles such as drones and rovers: physical-invariant based defenses~\cite{savior, ci}. Specifically, we select SAVIOR~\cite{savior} as a representative design since it adopts more principled state estimation models and thus shows superior detection performance over prior designs such as CI~\cite{ci}. The detailed setup for SAVIOR evaluation can be found in Appendix~\ref{app:savior_setup}.

%In our evaluation, we focus on SAVIOR~\cite{savior} rather than CI~\cite{ci} since SAVIOR adopts more complex nonlinear state estimation models and has shown superior detection performance than CI~\cite{savior}. 
\cut{
To evaluate SAVIOR, we follow the similar methodology as the ground rover evaluation in the SAVIOR paper~\cite{savior}, i.e., using the kinematic bicycle model~\cite{kong2015kinematic} and an Extended Kalman Filter (EKF) to predict the system state (i.e., position in x, y coordinates) given the vehicle control commands (i.e., steering and acceleration). Although the vanilla bicycle model does not have tunable parameters, we follow a similar implementation as SAVIOR by adding coefficients to the bicycle model equations~\cite{savior_code}. Same as SAVIOR, we use the \textit{nlgreyest} system identification tool from Matlab~\cite{matlab_si} to find the coefficients that can best fit the sensor and control trace.
During the evaluation, we continuously calculate the residuals between the GPS measurements and the predicted positions from the EKF, and feed the residuals to a CUSUM anomaly detector for attack detection. An execution that triggers the CUSUM detector will be considered as under attack.

Since the KAIST traces do not contain control outputs, we replay the traces in Baidu Apollo to record the vehicle control commands (more details in Appendix~\ref{app:savior_collect_control}). In addition to the KAIST traces, we also evaluate SAVIOR on a dataset that contains the original control commands to validate SAVIOR's detection performance in an ideal setting. However, similar performance is observed in that dataset to the ones on KAIST traces. More details of this are in Appendix~\ref{app:savior_with_real_control}.
}

\textbf{Evaluation metrics.}
As \ld{} involves two defense stages with different defense goals, we separate the evaluation into attack detection and response evaluations. For attack detection evaluate, we plot the \textit{ROC curves} to systematically show the TPRs and FPRs under different CUSUM parameters $b$ and $\tau$ (\S\ref{sec:design_detection}). In addition to ROC curves, we also report the maximum MSF lateral deviation before the attack is detected by \ld{}. This \textit{detection deviation} is a metric to indicate the detection timeliness, e.g., a detection deviation smaller than the lane straddling deviation (i.e., deviation to touch the lane line) means that the attack is detected early in time before it can cause any meaningful adversarial consequences. For attack response evaluation, we focus on the lateral deviations since our AR goal is to steer the vehicle to stop within the lane boundaries. In particular, we report two lateral deviation metrics with one measuring the \textit{maximum deviation} before the vehicle fully stops, and another one measuring the final \textit{stopping deviation}. In practice, the latter is more important since it 
% the former deviation is only temporal, but the latter
will be the permanent deviation after the vehicle stops.

\vspace{-0.05in}
\nsubsection{Attack Detection Effectiveness} \label{sec:eval_detection}
\vspace{0.03in}

\begin{figure*}[tbp]
\centering
\includegraphics[width=\linewidth]{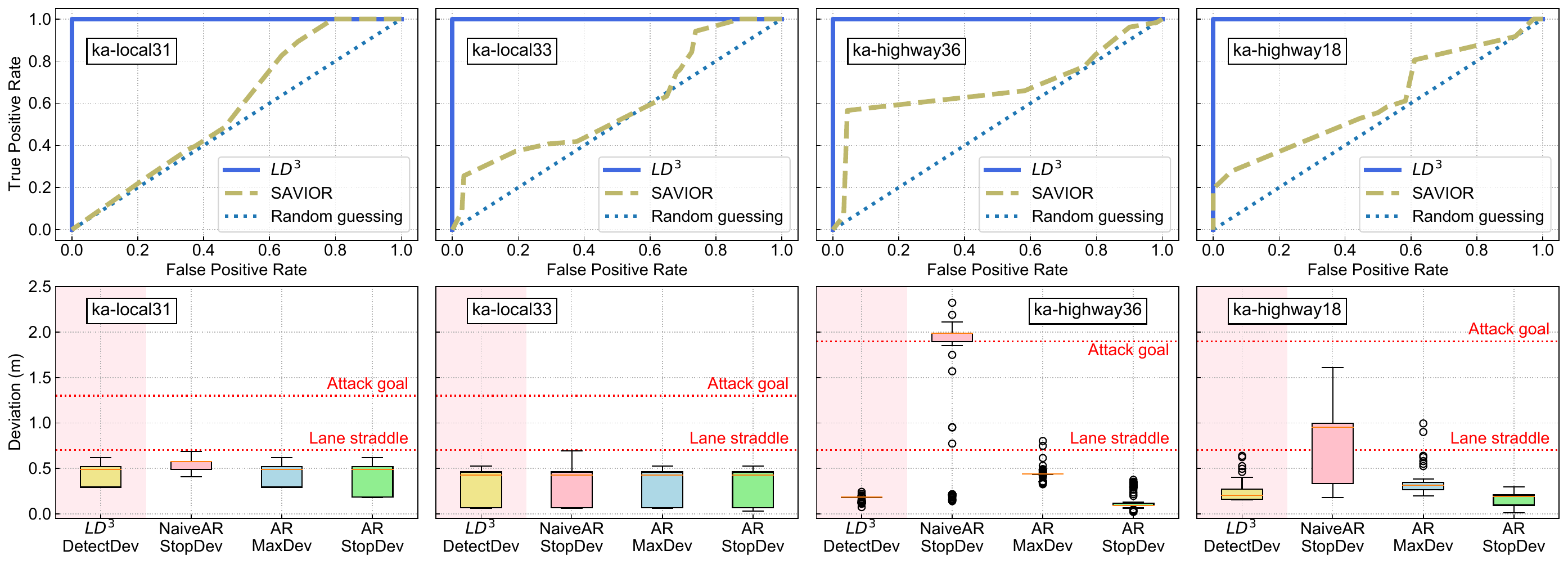}
% \vspace{-0.3in}
\caption{(Top) Attack detection ROC curves; (Bottom) Detection and Attack Response (AR) deviations in the \ld{} evaluation.}
\label{fig:trace_rocs_devs}
\vspace{-0.1in}
\end{figure*}

\textbf{Attack detection rates.}
The top figures in Fig.~\ref{fig:trace_rocs_devs} show the detection ROC curves of \ld{} against \fr{}. As shown, \ld{} can achieve effective detection with 100\% TPRs and 0\% FPRs on all 4 traces. During the search for best CUSUM parameters, we find that in benign drivings, differences between MSF and LD lateral deviations are always bounded within certain range ($<$0.6 m). However, in attacked drivings, \fr{} will cause larger lateral deviations on MSF side, which will be reflected on LD side in the opposite direction. Such a difference between benign and attacked drivings makes the attack easily detectable by \ld{}. Fig.~\ref{fig:cusum} shows an example of the benign and attacked MSF and LD lateral deviations and their CUSUM statistics. In the attacked case, \fr{} launches the vulnerability profiling stage from $t$=1544686730 and discovers a vulnerable window at $t$=1544686765, where MSF starts to exhibit larger lateral deviations. Because of the distinctive MSF/LD consistency levels between the benign and attack cases, it is thus straightforward to set a CUSUM threshold to differentiate them.

\textbf{Baseline comparison.} As shown, SAVIOR's detection performance is only slightly better than random guessing and far from being an ideal detector. Such a poor detection performance would render SAVIOR unpractical since it will introduce lots of false positives in normal driving.

The reason behind the poor detection performance in the AD context is twofold. First, compared to drones and rovers, the physical dynamics of the vehicle are much harder to model due to the complex physical moving characteristics, e.g., tire-road frictions, aerodynamic forces, road bank angles, etc.~\cite{rajamani2011vehicle}. For example, prior study~\cite{polack2017kinematic} finds that the error of kinematic bicycle model increases very fast at high speeds (e.g., 25 $\mathrm{m/s}$) or on curvy roads (e.g., steering angle at 4$^\circ$). In comparison, the bicycle model used in SAVIOR is reported having an average position error of 0.33 m \textit{within 0.8 sec} under low-speed settings (e.g., 13.8 m/s) in~\cite{kong2015kinematic} and its error keeps accumulating as time progresses; comparably, the same bicycle model incurs an average error of 1.076 m on \textit{ka-local31} within 1 sec, where the trace contains many turns and curvy roads.

Second, the attack deviation goals in the AD context can be much smaller but still being safety-critical. 
While SAVIOR is effective at detecting attacks on small robotics vehicles such as drones with large deviation goals (e.g., $\sim$50 m~\cite{savior}), attacks targeting high-level AD systems requires much smaller deviation and thus harder to detect. For example, even lateral deviations $<$0.5 m are enough to cause lane departure on narrow urban roads (e.g., 2.7 m wide~\cite{road_shoulder_width}).

\cut{
As discussed in \S\ref{sec:motivation_savior}, this is mainly because
the physical moving characteristics of AD vehicles are more complex and harder to accurately model than small robotics systems such as drones. Thus, the existing physical invariants (e.g., the bicycle model) used in SAVIOR fails to accurately estimate the system state. Another reason is that the attack goal deviations in the AD context are much smaller than the ones for drones, which makes the attack harder to detect.
% accurate enough to distinguish stealthy attacks such as \fr{}.
Such a distinctive detection performance between \ld{} and SAVIOR clearly shows the benefit of leveraging the lane boundary information source available on high-level AD systems.
}

\begin{table*}[tbp]
\footnotesize
\begin{minipage}{0.36\linewidth}
    \centering
    \includegraphics[width=\linewidth]{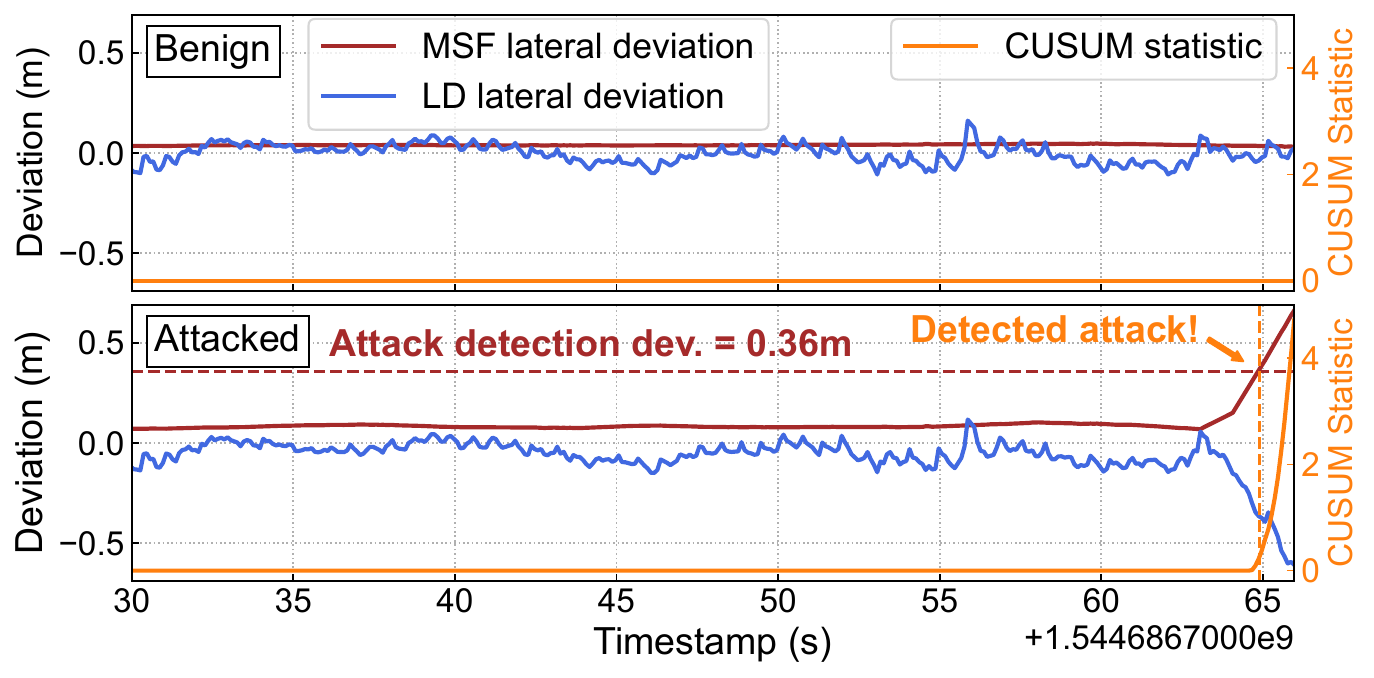}
    % \vspace{-0.3in}
    \captionof{figure}{Benign and attacked MSF/LD lateral deviations and CUSUM statistics.}
    \label{fig:cusum}
\end{minipage}\hfill
\begin{minipage}{0.36\linewidth}
    \centering
    \includegraphics[width=\linewidth]{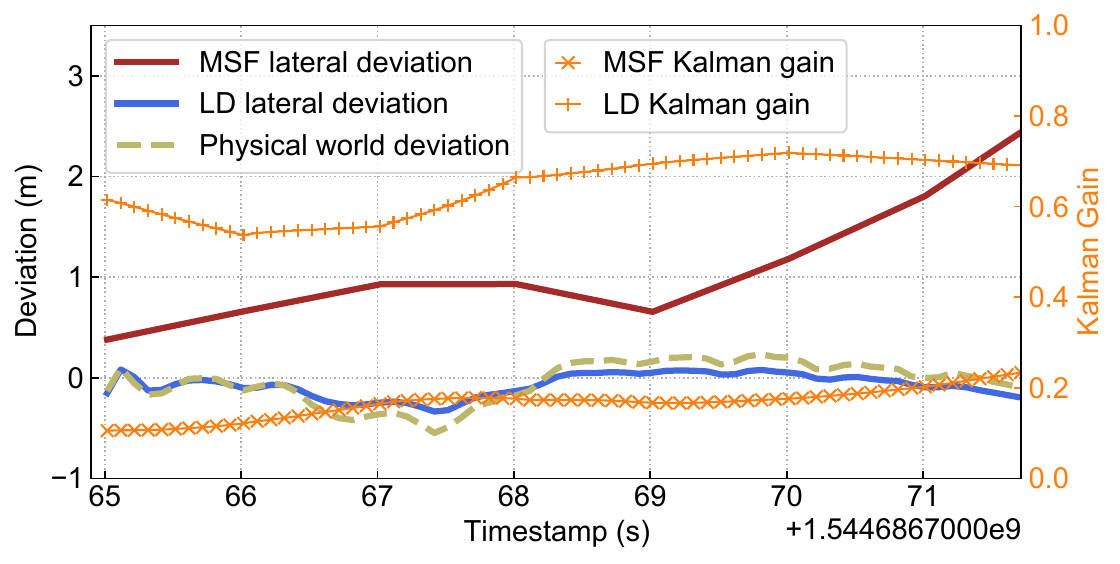}
    % \vspace{-0.3in}
    \captionof{figure}{MSF/LD and physical world deviations and Kalman gains during AR period.}
    \label{fig:ar_devs}
\end{minipage}\hfill
\begin{minipage}{0.245\linewidth}
    \centering
    \includegraphics[width=\linewidth]{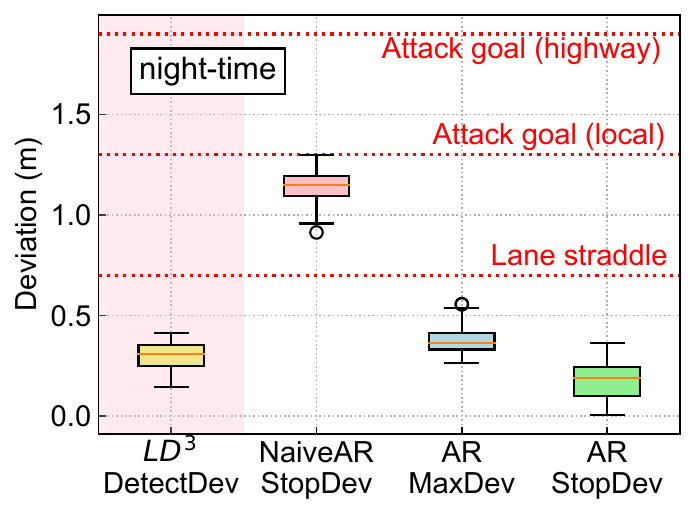}
    % \vspace{-0.3in}
    \captionof{figure}{Detection and AR deviations on the night-time trace.}
    \label{fig:real_car_devs}
\end{minipage}
\vspace{-0.2in}
\end{table*}

\textbf{Attack detection deviations.} To evaluate attack detection deviations, we choose a CUSUM weight $b=0.6$ and threshold $\tau=0.1$, which can achieve best attack detection effectiveness on all traces. For example, the detection deviation in Fig.~\ref{fig:cusum} is 0.36 m under these CUSUM parameters. The bottom figures in Fig.~\ref{fig:trace_rocs_devs} show the distributions of maximum deviations \fr{} has reached before being detected by \ld{} (box plots with pink background). As shown, \ld{} can promptly detect the attack before it can even cause lane straddling, and the average detection deviations are all below 0.5 m. However, there do exist two attack cases in \textit{ka-highway18} that the detection deviations are close to the lane straddling deviation. This is because in these attack cases, the lateral deviation at the MSF side raises very rapidly between detection intervals such that the deviation has already reached a large number (e.g., 0.63 m) before \ld{} has a chance to perform the detection. Nevertheless, none of the attack cases are detected after \fr{} starts to cause lane straddling and all of them are far away from reaching the attack goal deviation.
% (\textit{off-road} attack goal).
% This thus effectively reduces \fr{}'s attack success rates to 0\% on these traces.

\vspace{-0.05in}
\nsubsection{Attack Response Effectiveness} \label{sec:eval_ar}
\vspace{0.03in}

The distributions of the maximum deviations and final stopping deviations are shown in the bottom figures in Fig.~\ref{fig:trace_rocs_devs} (box plots without background colors). During the AR periods, none of the attack cases have a maximum or stopping deviation over the attack goal deviation (1.3 m for local and 1.9 m for highway). Despite 4 attack cases on the highway traces have maximum deviations exceed the lane straddling deviation (0.7 m), their stopping deviations are all corrected back to be within the lane boundaries.
% An interesting finding is that in many cases, the stopping deviations, especially on the highway traces, are even smaller than the attack detection deviations as shown in the figure.
This shows that our AR design (\S\ref{sec:design_ar}) is effective at keeping the vehicle within the lane boundaries when it stops, which can prevent the much more dangerous situation where it stops out of the ego lane.

Moreover, comparing between local and highway, the highway traces often have \textit{larger maximum deviations} and \textit{smaller stopping deviations}. This is because the driving speeds when the attacks are detected on the highway traces (27.3 $\mathrm{m/s}$ on avg.) are much higher than that on the local traces (3.8 $\mathrm{m/s}$ on avg.). This leads to a much longer AR period on highways ($\sim$7 sec on avg.) than that on local roads ($<$1 sec on avg.). As a result, \fr{} can keep causing larger lateral deviations after the attack is detected, but in the meanwhile, our AR design can also correct more given the longer AR period.

\cut{
\begin{figure}[tbp]
\centering
\includegraphics[width=.85\linewidth]{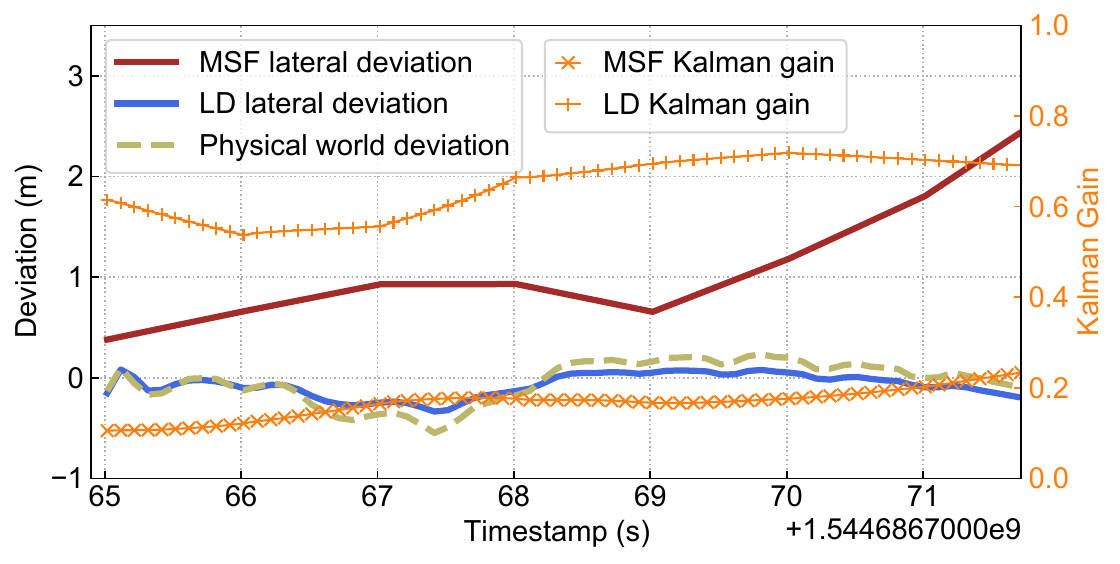}
\vspace{-0.15in}
\caption{MSF/LD and physical world deviations and Kalman gains during the attack response period (7 sec to stop).}
\label{fig:ar_devs}
\vspace{-0.05in}
\end{figure}
}

Fig.~\ref{fig:ar_devs} shows an example of the MSF/LD and physical world deviations during the AR period on \textit{ka-highway36}, where the maximum deviation and stopping deviation are 0.52 m and 0.19 m, respectively. In this example, since \fr{} keeps increasing the deviation on the MSF side, the safety-driven fusion (\S\ref{sec:design_ar}) penalizes the lateral deviations on the MSF side with higher uncertainties and thus results in smaller Kalman gains, which indicate the weights of the inputs in KF update. Consequently, the fusion process prioritizes the lateral deviations on the LD side, which are similar to the physical world deviations, and the lateral controller thus can steer the vehicle towards the right direction.

\textbf{Comparison with naive AR design.}
A naive AR design, named NaiveAR, applies the maximum deceleration to stop but still keeps using the MSF outputs for steering. Such design is similar to the \textit{in-lane stop} planning scenario that Baidu Apollo adopts to handle emergencies~\cite{apollo_planning}. To evaluate this, we record the MSF lateral deviations at the end of AR periods and regard them as the stopping deviations based on the control assumption (\S\ref{sec:background_threat_model}). The stopping deviations of NaiveAR are shown in Fig.~\ref{fig:trace_rocs_devs}. Because of the longer AR periods on the highway, the stopping deviations under NaiveAR are significantly higher than that using our complete AR design, especially on the highway traces. In particular, since the lateral deviations on \textit{ka-highway36} increase very quickly, over 75\% of the attacked cases still reaches a lateral deviation higher than the attack goal deviation, which consequently leads to $>$75\% attack success rate for \fr{} on \textit{ka-highway36} despite the attacks are correctly detected. On the other hand, with the complete AR design, none of the attack cases can be even deviate out of the lane boundaries.

% \begin{figure}[tbp]
% \centering
% \includegraphics[width=\linewidth]{./figs/test.png}
% \vspace{-0.1in}
% \caption{Test.}
% \label{fig:test}
% \end{figure}

\vspace{-0.05in}
\nsubsection{Evaluation under Limited Visibility} \label{sec:eval_visibility}
\vspace{0.05in}

% \junjie{consider removing the real-time overhead evaluation since we already conduct real-time vehicle evaluation with the defense.}

% Although the KAIST traces already include different time-of-day and road conditions (\S\ref{sec:eval_method}), in this section, we further evaluate \ld{} on a \textit{night-time} driving trace.
% that we collected using a vehicle with an Advanced Driver-Assistance System (ADAS).

\cut{  % revision: deleted since simulation & PIXKIT experiments are all real-time
\textbf{Real-time performance overhead evaluation setup.}
Among the \ld{} design components as shown in Fig.~\ref{fig:design_overview}, only the attack detection (\S\ref{sec:design_detection}) and the attack-aware fusion (\S\ref{sec:design_ar}) are on the timing critical path since they are performed as post-processing steps in the localization module. Compare between them, the attack detection logic is by design more time consuming (with more semantic map queries) and is the only one that will be executed in benign driving, thus our performance overhead evaluation focuses on this step. To do that, we implement the attack detection logic on an embedded ADAS device named EON~\cite{eon}, which is the official device to run OpenPilot~\cite{openpilot}. Since OpenPilot adopts a similar modular design with parallel processes such as localization, LD, etc. Thus, similar to Fig.~\ref{fig:design_overview}, we implement the attack detection as a post-processing step in the localization process and measure its average timing overhead over a 30-min duration.
}

\begin{figure}[tbp]
\centering
\includegraphics[width=\columnwidth]{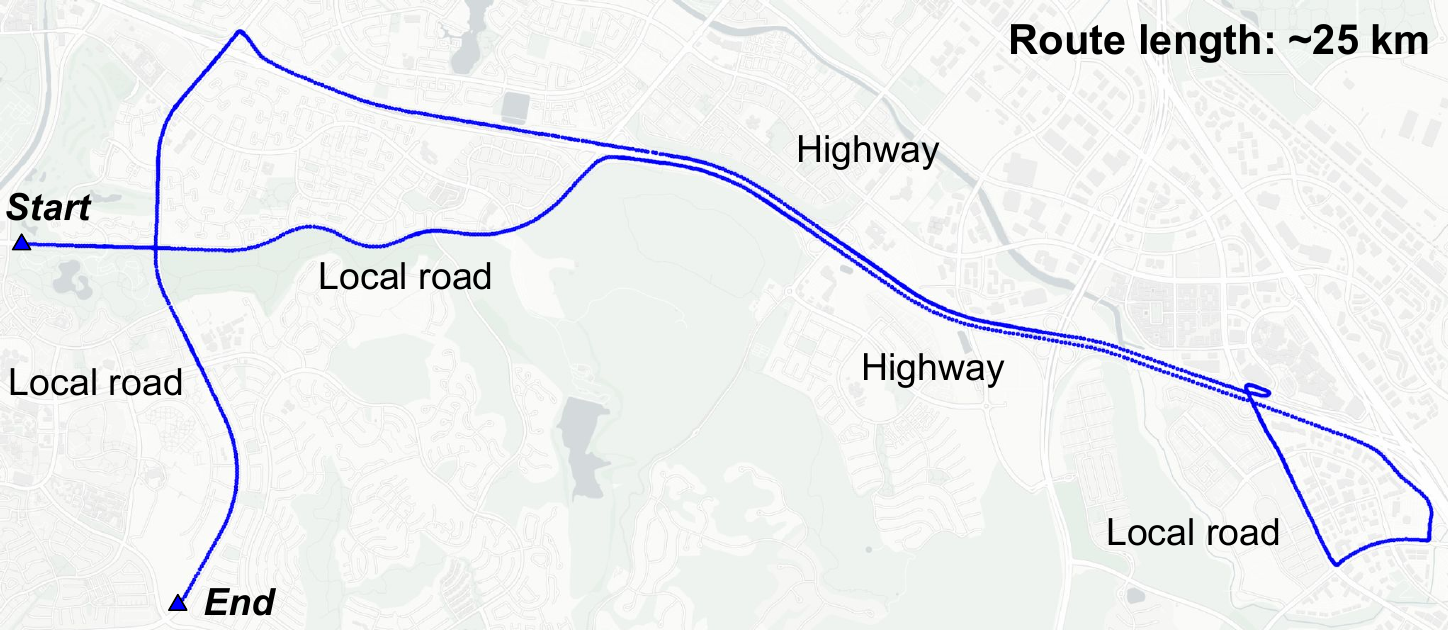}
% \vspace{-0.1in}
\caption{Route of the night-time sensor trace collected using EON~\cite{eon} in \S\ref{sec:eval_visibility}.}
\label{fig:real_car_map}
% \vspace{-0.05in}
\end{figure}

\textbf{Trace collection and defense evaluation setup.} We collect a \textit{night-time} driving trace at around 11 p.m. our local time using an Advanced Driver-Assistance System (ADAS) device named EON~\cite{eon}, which is the official device to run OpenPilot~\cite{openpilot}. Specifically, we record the localization and LD outputs during the trace collection for the defense evaluation.
% EON can provide the necessary inputs (localization and LD) for the defense evaluation, we thus use it to collect a real-world driving trace and conduct a trace-based evaluation similar to \S\ref{sec:eval}. 
% Specifically, we install the EON device on a Toyota Camry 2019 and record the localization and LD outputs during the trace collection.
% Note that during the trace collection, the vehicle is manually driven by us rather than by OpenPilot since it cannot be constantly engaged to control the vehicle due to the intersections, which are out of the applicable domain for LD. 
The trace is $\sim$25 km in length with 3 local road and 2 highway segments as shown in Fig.~\ref{fig:real_car_map}. 
% The average speeds in the trace range from 10.3 to 27.5 $\mathrm{m/s}$ due to the dynamic traffic and different speed limits. 
% To evaluate \ld{}, we follow the same methodology as in Appendix~\ref{app:map_creation} to create the semantic map.
Since EON does not provide LiDAR data, we are not able to run MSF and \fr{} attack. To model the attack effect, we apply the lateral deviations from the \textit{most aggressive} attack trace in \textit{ba-local} trace used in the \fr{} paper~\cite{fusionripper} to the localization outputs, which only takes 10 sec from the start of attack to reaching a 2 m lateral deviation. This is similar to the prior works where they directly apply the attack traces in the target systems for attack detection evaluation~\cite{savior, ci}. Specifically, we apply the attack trace consecutively to all road segments excluding the intersections, which results in 98 attacked and 98 corresponding benign segments in total.

% u-blox GNSS receiver uses multi-GNSS~\cite{ublox_multignss} (e.g., GPS \& GLONASS) to provide more accurate positioning.

% \begin{figure}[tbp]
% \centering
% \includegraphics[width=0.9\columnwidth]{./figs/real_car_map.pdf}
% \vspace{-0.1in}
% \caption{Trace collected using our own vehicle. The average speed is annotated for each local road and highway segment.}
% \label{fig:real_car_map}
% \vspace{-0.03in}
% \end{figure}

\cut{  % revision: deleted since simulation & PIXKIT experiments are all real-time
\textbf{Real-time performance overhead.}
During the 30-min period, we find that the attack detection logic in \ld{} only takes 0.37 ms on average (std: 0.26 ms). In comparison, LD runs at 20 Hz in the EON, which means that each LD output can tolerate a 50 ms processing delay. Thus, the timing overhead of the attack detection in \ld{} is negligible even on an embedded ADAS device such as EON. Commercial high-level AD systems often run on Industrial PCs~\cite{apollo_hardware}, which are much more powerful and thus the overhead might be even lower.
}

\textbf{Defense effectiveness.}
Similar to results on KAIST traces (\S\ref{sec:eval_detection}), \ld{} can achieve effective attack detection with 100\% TPR and 0\% FPR on the night-time trace. The attack detection and AR deviations are shown in Fig.~\ref{fig:real_car_devs}. As shown, even under such low-light condition, \ld{} can still timely detect the attack with an average detection deviation of 0.29 m. Consistent with findings in \S\ref{sec:eval_ar}, the stopping deviation on the real vehicle trace is only 0.17 m on average, which means \ld{} is effective at stopping the vehicle within the lane boundaries. In comparison, NaiveAR has a stopping deviation much higher than \ld{}, where one attack segment (maximum deviation is 1.33 m) exceeds the goal deviation for local roads. 

% detection: mean, std: 0.31 0.07
% EH max: mean, std: 0.36 0.08
% EH stop: mean, std: 0.13 0.10

\cut{
\begin{figure}[tbp]
\centering
\includegraphics[width=.5\columnwidth]{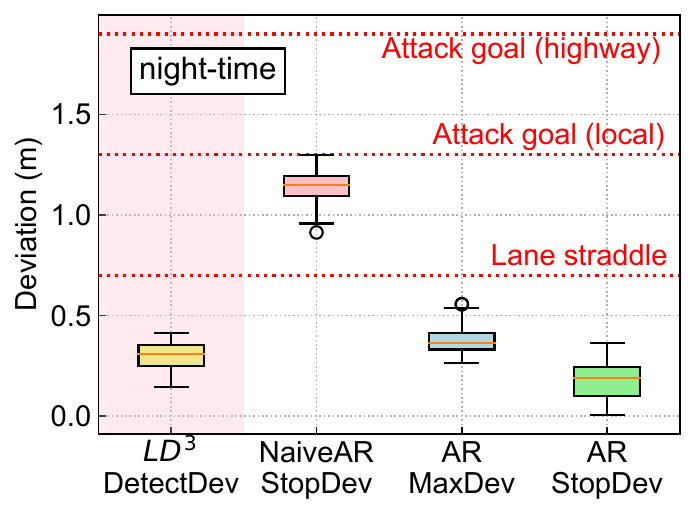}
\vspace{-0.1in}
\caption{\ld{} detection and AR deviations in the defense evaluation on the night-time sensor trace. \ld{} can achieve 100\% TPR and 0\% FPR in attack detection on this trace.}
\label{fig:real_car_devs}
\vspace{-0.05in}
\end{figure}
}
% \vspace{-0.05in}
\nsection{End-to-End Evaluations} \label{sec:end_to_end_eval} 
% \vspace{-0.05in}

% \newparts{
% In this section, we implement \ld{} on 2 open-source full-stack AD systems, Baidu Apollo~\cite{apollo} and Autoware~\cite{autoware}, and evaluate its defense capability in end-to-end driving with closed-loop control in an industry-grade high-level AD simulator and on a real vehicle-sized AD development chassis. The demo videos are available on our project website at 
% \textbf{\url{https://sites.google.com/view/ld3-defense}}.}

\newparts{
In this section, we implement \ld{} on 2 open-source full-stack AD systems, Baidu Apollo~\cite{apollo} and Autoware~\cite{autoware}, and evaluate \ld{} under end-to-end drivings in both simulation and the physical world. The demo videos are available on our project website at 
\textbf{\url{https://sites.google.com/view/cav-sec/LD3}}.}

% integrate \ld{} in an industry-grade high-level AD system and show the defense effectiveness in end-to-end simulation environments with the presence of AD control.

\vspace{-0.05in}
\nsubsection{Evaluation in AD Simulator}  \label{sec:simulation}

\cut{
\textbf{Experimental setup.}
We implement \ld{} in Baidu Apollo v5.0.0~\cite{apollo} and evaluate under 4 driving scenarios with different driving speeds (local road and highway speeds) and road geometries (straight and curvy roads) in the LGSVL simulator~\cite{lgsvl}.
% following the design in Fig.~\ref{fig:design_overview}. We run the complete Baidu Apollo AD system with all functional modules enabled in an industry-grade AD simulator, LGSVL~\cite{lgsvl}. 
% We evaluate the benign and attacked drivings with \ld{} in 4 driving scenarios with different driving speeds (local road and highway speeds) and road geometries (straight and curvy roads).
In our evaluation, we include the \ld{} variant with naive AR design (\S\ref{sec:eval_ar}) and one without defense.
We repeat the simulation for 10 times with different attack starting times for each combination of simulation scenarios and defense settings.
The detailed simulation setup is in Appendix~\ref{app:simulation_setup}.}

\textbf{Experimental setup.}
We implement \ld{} in Baidu Apollo v5.0.0~\cite{apollo} following the design in Fig.~\ref{fig:design_overview}. Specifically, we reuse the SCNN model~\cite{pan2018spatial} for LD, which is currently used only for camera calibration in Baidu Apollo. We run the complete Baidu Apollo AD system with all functional modules enabled in a production-grade AD simulator, LGSVL~\cite{lgsvl}. Since LGSVL does not provide LiDAR locator maps required for MSF, we instead run Baidu Apollo localization in the Real-Time Kinematic mode, which directly takes the ground truth positions from LGSVL. To simulate the \fr{} attack effect, we add the lateral deviations from the same attack trace used in \S\ref{sec:eval_visibility} to the localization outputs.

We evaluate the benign and attacked drivings with \ld{} in 4 driving scenarios on two LGSVL maps: Single Lane Road (SLR) and San Francisco (SF). Specifically, the SLR map is a long straight road, and we create a low-speed (SLR-Low) and high-speed (SLR-High) driving scenario on it by adjusting the maximum cruising speed in Apollo planning. The SF map is a 1:1 re-creation of a portion of the San Francisco city, from which we select a straight (SF-Straight) and a curvy road (SF-Curvy). In our evaluation, we also include the \ld{} variant with the naive AR design (\S\ref{sec:eval_ar}) and a setting without any defenses. We repeat the simulation for 10 times with different attack starting times for each combination of simulation scenarios and defense settings.

\begin{figure}[tbp]
\centering
\includegraphics[width=.95\linewidth]{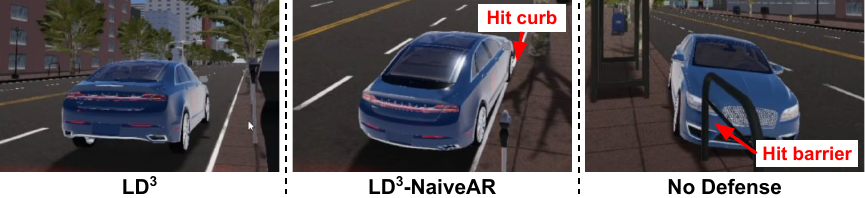}
% \vspace{-0.15in}
\caption{Simulation snapshots of the vehicle stopping locations under the 3 defense settings in SF-Straight.}
\label{fig:sim_snapshot}
\vspace{-0.05in}
\end{figure}

\textbf{Results and demos.}
Our simulation results show that the attack detection rates for both \ld{} and \ld{}-NaiveAR are all 100\% in the 10 runs, and none of the benign drivings are falsely detected as under attack. 
Table~\ref{tbl:sim_results} shows the maximum lateral deviation achieved in the whole simulation (including both attack detection and response periods) in each scenario/defense setting and the corresponding vehicle stopping location. As shown, with \ld{}, the average maximum deviations are smaller than lane straddling deviation in all 4 scenarios and the vehicle can always safely stop in the lane. In comparison, due to the blind trust of the localization outputs in the AR period, \ld{}-NaiveAR has much higher maximum deviations than \ld{} and the vehicle's stopping locations are either lane straddling or already crashing into the road curb/barrier. Nevertheless, the No Defense setting is even worse than \ld{}-NaiveAR, where the vehicle is simply deviated to fall off the road in SLR-Low and SLR-High.
Snapshots of the vehicle stopping locations in SF-Straight are shown in Fig.~\ref{fig:sim_snapshot}. 
The demos of the 4 simulation scenarios and 3 defense settings are available on our project website.

% \textbf{\url{https://sites.google.com/view/ld3-defense}}

\begin{table*}[tbp]
\footnotesize
\centering
\caption{Maximum deviations to lane center and attack consequences under different defense settings in the 4 simulation scenarios in \S\ref{sec:simulation}. Each setting was run for 10 times with randomized attack starting times. Benign driving with \ld{} is also presented and was run for 10 times. The maximum deviations are represented as (mean, std) in meters.}
\label{tbl:sim_results}
% \vspace{-0.1in}
\setlength{\tabcolsep}{4.5pt}
\begin{tabular}{@{}c|c|cccccc|cc@{}}
\toprule
\multirow{3}{*}{\begin{tabular}[c]{@{}c@{}}Simulation\\ scenario\end{tabular}} & \multirow{3}{*}{\begin{tabular}[c]{@{}c@{}}Lane\\ straddle\\ dev\end{tabular}} & \multicolumn{6}{c}{Attacked} & \multicolumn{2}{|c}{Benign} \\ \cmidrule(l){3-10} 
 &  & \multicolumn{2}{c}{\ld} & \multicolumn{2}{c}{\ld-NaiveAR} & \multicolumn{2}{c}{No Defense} & \multicolumn{2}{|c}{\ld} \\ \cmidrule(l){3-10} 
 &  & Max dev & Consequence & Max dev & Consequence & Max dev & Consequence & Max dev & Consequence \\ \midrule
SLR-Low & 0.83 & 0.47, 0.08 & Stop in lane & 1.69, 0.06 & Stop w/ lane straddle & 7.94, 0.05 & Fall off road & 0.07, 5e-5 & Reach destination \\
SLR-High & 0.83 & 0.69, 0.06 & Stop in lane & 1.64, 0.16 & Stop w/ lane straddle & 7.93, 0.04 & Fall off road & 0.07, 5e-5 & Reach destination \\
SF-Straight & 1.00 & 0.67, 0.23 & Stop in lane & 1.02, 0.01 & Hit curb & 1.84, 0.16 & Hit tree or barrier & 0.14, 7e-4 & Reach destination \\
SF-Curvy & 0.75 & 0.43, 0.14 & Stop in lane & 0.90, 0.12 & Hit lane divider & 0.97, 0.14 & Hit lane divider & 0.31, 0.01 & Reach destination \\ \bottomrule
\end{tabular}
% \vspace{-0.1in}
\end{table*}

% \vspace{0.05in}
\nsubsection{\newparts{Evaluation on AD Development Chassis of Real Vehicle Size and Closed-loop Control}} \label{sec:pixkit_eval}
\vspace{0.03in}

% Although existing AD simulation technologies are already widely used in AD development for safety testing, it is still unclear whether the physics modeling is accurate enough such that the defense capability can indeed be faithfully translated to real-world driving. Therefore, we further perform another end-to-end evaluate of \ld{} on a real vehicle-sized AD development chassis with closed-loop control.
% \newparts{In this section, we evaluate \ld{} on a real vehicle-sized AD development chassis with closed-loop control to understand the defense capabilities in the real world.}

\begin{figure}[tbp]
\centering
\includegraphics[width=\columnwidth]{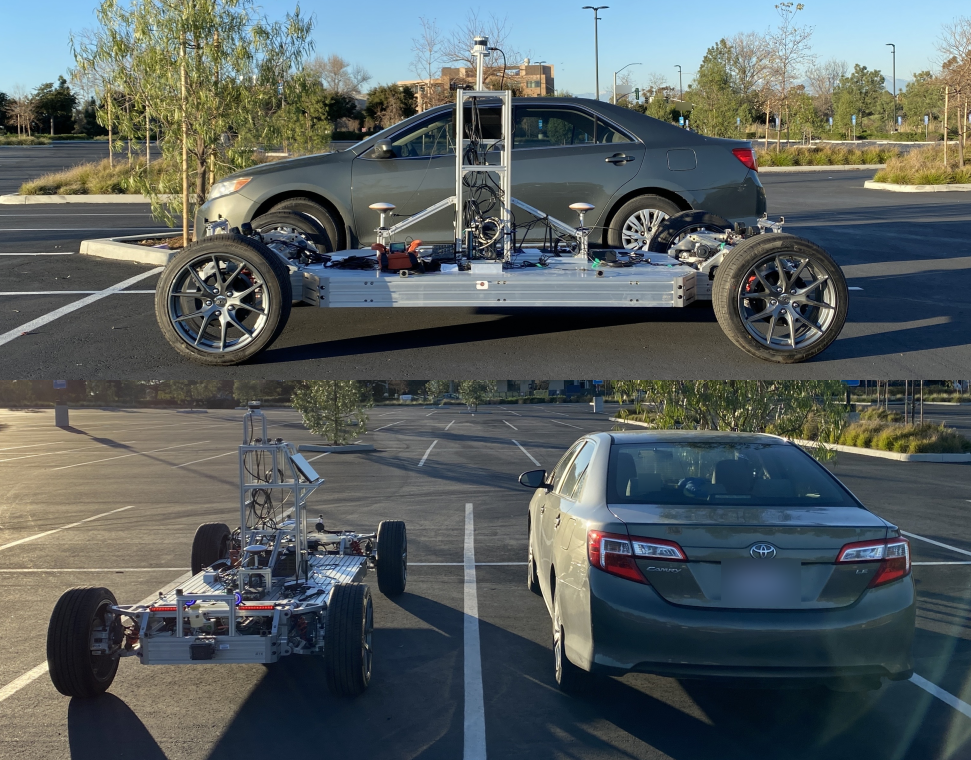}
% \vspace{-0.1in}
\caption{Side-by-side views of the AD development chassis used in \S\ref{sec:pixkit_eval} and a Toyota Camry.}
\label{fig:pixkit_side_by_side}
% \vspace{-0.05in}
\end{figure}

\textbf{Experimental setup.} We experiment on an AD chassis as shown in Fig.~\ref{fig:pixkit_side_by_side}, which is specifically designed for Level-4 AD system prototyping and testing. The chassis is of a real vehicle size, capable of closed-loop control, and fully equipped with Level-4 AD sensors including LiDAR, GPS, IMU, cameras, RADARs, and ultrasonic sensors.
Since AD vehicle testing is not allowed to be on public roads by default, we reserve a parking lot in our institute for the experiments. Specifically, we mark a straight traffic lane with 3.5 m width (the most common lane width in KAIST dataset and our night-time driving trace) in the parking lot and create the corresponding semantic map for Autoware.

We ported \ld{} to the Autoware AD system~\cite{autoware}, which is currently supported by the AD chassis. To facilitate the attack, we apply the same \fr{} attack trace used in \S\ref{sec:eval_visibility} and \S\ref{sec:simulation} to the localization outputs in Autoware.
Unlike OpenPilot and Baidu Apollo, the lane detector in Autoware can only detect lane lines in pixels rather than in the world coordinates. Therefore, we directly obtain the ground truth lane line information from the map using the unmodified localization outputs, since LD is already a mature technology (\S\ref{sec:opportunity}) and has been shown to be quite accurate in \S\ref{sec:eval} and \S\ref{sec:simulation}.
We enable the relevant components in Autoware including localization, global/local plannings, and control. During the experiments, the AD chassis is completely driven by Autoware unless taken over by us from a remote controller in emergency situations. We evaluate three defense settings: (1) \textit{w/ \ldi{} w/ attack}, (2) \textit{w/o \ldi{} w/ attack}, and (3) \textit{w/ \ldi{} w/o attack}.
For each, we experiment in driving speeds of 2 m/s (4.5 mph) and 4 m/s (9 mph) for safety concerns. 
We prolong the AR stage by using deceleration $<$3 $m/s^2$ in both cases to better showcase the driving behaviors during AR.
Specifically, we repeat the experiments for 3 times for \textit{w/ \ldi{} w/ attack}. Since the other two are always quite stable, we thus do not record more iterations for those experiments.

\begin{table*}[tbp]
\footnotesize
\begin{minipage}{0.55\linewidth}
    \footnotesize
    \centering
    \caption{The detection, maximum, and stopping deviations in the three settings at two different driving speeds. We repeat the experiments for \textit{w/ \ld{} w/ attack} for 3 times and report the (mean, std) deviations. We do not repeat the other two settings as they are quite stable.}
    \label{tbl:pixkit_devs}
    % \vspace{-0.1in}
    \setlength{\tabcolsep}{1.2pt}
    \begin{tabular}{@{}c|cccc|cc@{}}
    \toprule
    \multirow{3}{*}{Speed} & \multicolumn{4}{c|}{w/ attack} & \multicolumn{2}{c}{w/o attack} \\ \cmidrule(l){2-7} 
     & \multicolumn{3}{c|}{w/ \ld{}} & w/o \ld{} & \multicolumn{2}{c}{w/ \ld{}} \\ \cmidrule(l){2-7} 
     & Det dev & Max dev & \multicolumn{1}{c|}{Stop dev} & Max/Stop dev & Max dev & Stop dev \\ \midrule
    4 m/s & 0.07m, 0.01m & 0.36m, 3e-3m & \multicolumn{1}{c|}{0.05m, 0.05m} & 2.59m & 0.13m & 8e-3m \\
    2 m/s & 0.02m, 2e-3m & 0.27m, 0.04m & \multicolumn{1}{c|}{0.01m, 1e-3m} & 2.23m & 0.11m & 7e-3m \\ \bottomrule
    \end{tabular}
\end{minipage}\hfill
\begin{minipage}{0.43\linewidth}
    \footnotesize
    \centering
    \caption{Maximum physical deviations can be achieved without being detected under various LD fluctuation assumptions. The percentages indicate the probabilities of such fluctuations.}
    \vspace{-0.05in}
    \label{tbl:stealthy_detection}
    \setlength{\tabcolsep}{3pt}
    \begin{tabular}{@{}ccccc@{}}
    \toprule
    \multirow{2}{*}{Trace} & \multirow{2}{*}{\begin{tabular}[c]{@{}c@{}}LD fluctuation\\ ($\mu, \sigma$)\end{tabular}} & \multicolumn{3}{c}{Max physical world deviation} \\ \cmidrule(l){3-5} 
     &  & 0 (100\%) & $\mu$ (50\%) & $\mu+3\sigma$ (0.3\%) \\ \midrule
    \textit{ka-local31} & 0.12m, 0.08m & 0.7m & 0.82m & 1.06m \\
    \textit{ka-local33} & 0.14m, 0.10m & 0.7m & 0.84m & 1.14m \\
    \textit{ka-highway36} & 0.29m, 0.10m & 0.7m & 0.99m & 1.29m \\
    \textit{ka-highway18} & 0.20m, 0.11m & 0.7m & 0.90m & 1.23m \\ \bottomrule
    \end{tabular}
\end{minipage}
% \vspace{-0.22in}
\end{table*}

% 4 m/s:
% Detection dev (mean, std): 0.06493076049897435 0.012452686654347397
% Maximum dev (mean, std): 0.36426716650795504 0.0032264096321182583
% Stopping dev (mean, std): 0.04520309618108121 0.04720487829962962
% NoDefense Maximum dev: 2.5913297119226946
% NoDefense Stopping dev: 2.584318858073847
% Benign Maximum dev: 0.13370482650200852
% Benign Stopping dev: 0.00814336299953038

% 2 m/s:
% Detection dev (mean, std): 0.020136245340817385 0.0022321999365849127
% Maximum dev (mean, std): 0.2731832984655125 0.04088672062665852
% Stopping dev (mean, std): 0.0058209895419398605 0.0017981851847734464
% NoDefense Maximum dev: 2.2288639764466422
% NoDefense Stopping dev: 2.2078183052640297
% Benign Maximum dev: 0.10658935648516672
% Benign Stopping dev: 0.00735364768826751

\textbf{Results and demos.} 
Table~\ref{tbl:pixkit_devs} shows the detection, maximum, and stopping deviations under the three settings. As shown, \ld{} on average can detect the attack when the vehicle's physical deviation is still small and start the AR stage. Within the AR period, the average maximum deviations are 0.36 m and 0.27 m at speeds of 4 m/s and 2 m/s, respectively, and the final stopping deviations are always within 0.1 m. In comparison, without \ld{}, the vehicle keeps deviating and we have to manually press the emergency button on the remote to prevent it from crashing into the curb. 
Such a distinctive driving behaviors with and without \ld{} are consistent with our trace-based (\S\ref{sec:eval}) and simulation results (\S\ref{sec:simulation}).
Without the attack, the vehicle's trajectories well align with the road centerline (i.e., the reference trajectory Autoware plans to enforce) and eventually complete the route and stop at the center of the lane.
We also record demo videos of the vehicle driving behaviors under the three settings (videos are available on our website). As an illustration, Fig.~\ref{fig:pixkit_stop_positions} visualizes the driving trajectories in the bird's eye view and shows the snapshots of final stopping positions at driving speed of 4 m/s.

% \todo{put the PIXKIT evaluation here as a subsection.
% 2 m/s: maximum deviation in AR: 0.23 meters, final stop deviation: 0.09 meters
% 4 m/s: maximum deviation in AR: 0.37 meters, final stop deviation: 0.10 meters}

% \vspace{-0.05in}
\nsection{Evaluation against Adaptive Attacks} \label{sec:adaptive_attacks}

% In \S\ref{sec:eval} and \S\ref{sec:end_to_end_eval}, we demonstrate the effectiveness of \ld{} at detecting and responding to existing state-of-the-art lateral-direction localization attack. 
In this section, we take a step further to examine \ld{}'s capability under potential adaptive attacks, including (1) an idealized stealthy attack that can evade the detection, and (2) the latest LD-side attack, which is the inherent new attack surface introduced by \ld{} approach (\S\ref{sec:design_challenges}).

% \vspace{-0.05in}
\nsubsection{Stealthy Attack Evaluation} \label{sec:stealthy_attack}
\vspace{0.05in}

% Since the evaluation in the above sections are all based on \fr{}, which does not assume the existence of defenses such as \ld{}. However, \fr{} is a realistic evaluation target since it is by far the only attack that can break the MSF localization on high-level AD systems. Nevertheless, 
In this evaluation, we analyze the maximum lateral deviations that a hypothetical stealthy attack can achieve by assuming stronger and unrealistic attack capabilities.

% assume the attacker can have \textit{precise} and \textit{instant} control over the lateral deviations in the MSF localization outputs, and evaluate the maximum lateral deviations a hypothetical stealthy attack can cause without being detected.

\textbf{Evaluation methodology.}
Based on the CUSUM anomaly detection formulation (\S\ref{sec:design_detection}), the attack should satisfy $S_{i-1} + \lvert D_i^{\text{MSF}} - D_i^{\text{LD}} \rvert - b < \tau$ in order to prevent detection. Assuming the last CUSUM statistic $S_{i-1} = 0$, the maximum MSF lateral deviation without being detected is thus $D_{i, max}^{\text{MSF}} = D_i^{\text{LD}} + \tau + b$, which is also the maximum physical world deviation given the control assumption (\S\ref{sec:background_threat_model}).
% As shown, the deviation that a stealthy attack can cause depends on the CUSUM parameters and the LD lateral deviation.
Since $\tau$ and $b$ are fixed in the defense, the attacker can carefully select a timing where the LD has a large lateral deviation fluctuation to the actual vehicle location due to detection noises, and apply the MSF lateral deviation to the same direction as the LD's fluctuation direction to achieve a large physical world deviation. Therefore, the attacker's capability on capturing a particular LD fluctuation window determines the maximum physical world deviations she can achieve without being detected. Thus, we evaluate the maximum physical world deviations by assuming various levels of LD fluctuations that the attacker can capture.

% that the attacker can capture particular levels of LD fluctuations. 

\textbf{Assumptions on attack capabilities.}
In this evaluation, we assume the attacker has very unrealistic attack capabilities in order to achieve such a stealthy attack. In particular, the attacker should have a \textit{white-box knowledge} on (1) where exactly on the road that the LD will have a large fluctuation and how much it is, and (2) the attack detection method and parameters used in the target AD system. Moreover, the attacker should also have \textit{precise} and \textit{instantaneous} control over the lateral deviations in the MSF localization outputs in order to execute such attack when large fluctuations appear.

\cut{
\begin{table}[tbp]
\centering
\footnotesize
\caption{Maximum physical deviations can be achieved without being detected under various LD fluctuation assumptions. The percentages indicate the probabilities of such fluctuations.}
\vspace{-0.1in}
\label{tbl:stealthy_detection}
\begin{tabular}{@{}ccccc@{}}
\toprule
\multirow{2}{*}{Trace} & \multirow{2}{*}{\begin{tabular}[c]{@{}c@{}}LD fluctuation\\ ($\mu, \sigma$)\end{tabular}} & \multicolumn{3}{c}{Max physical world deviation} \\ \cmidrule(l){3-5} 
 &  & 0 (100\%) & $\mu$ (50\%) & $\mu+3\sigma$ (0.3\%) \\ \midrule
\textit{ka-local31} & 0.12m, 0.08m & 0.7m & 0.82m & 1.06m \\
\textit{ka-local33} & 0.14m, 0.10m & 0.7m & 0.84m & 1.14m \\
\textit{ka-highway36} & 0.29m, 0.10m & 0.7m & 0.99m & 1.29m \\
\textit{ka-highway18} & 0.20m, 0.11m & 0.7m & 0.90m & 1.23m \\ \bottomrule
\end{tabular}
\vspace{-0.05in}
\end{table}
}

\textbf{Results.}
Table~\ref{tbl:stealthy_detection} shows the maximum physical world deviations that the stealthy attack can achieve under different LD fluctuation assumptions. Specifically, we calculate LD fluctuation distributions in each trace and assume that the attacker knows where a certain level of fluctuation happens. Without any such assumptions, the attacker can at most inject $\tau+b=0.7$ m lateral deviation, which is just about to touch the lane boundaries. On the other hand, the attacker can \textit{at most} cause 0.99 m and 1.29 m lateral deviations on the 4 traces if she can capture an average and a 3-$\sigma$ LD fluctuation, respectively. Note that the probabilities of such fluctuations to appear are 50\% and 0.3\% according to the normal distribution. In conclusion, even under very unrealistic attack assumptions, the maximum lateral deviations are still less than the local road attack goal (1.3 m) for \fr{}, which shows that \ld{} is quite effective at bounding the lateral deviations. Moreover, it also highlights that LD is indeed a mature technology (\S\ref{sec:opportunity}) suitable for defense given its high stability.
% means that LD is a reliable defense source due to its stability. 

% \begin{table}[tbp]
% \centering
% \footnotesize
% \caption{Maximum deviations that can be achieved without being detected by \ld.}
% \label{tbl:stealthy_detection}
% \setlength{\tabcolsep}{2pt}
% \begin{tabular}{@{}cccc@{}}
% \toprule
% Trace & \begin{tabular}[c]{@{}c@{}}Max Dev w/o\\ Detection\end{tabular} & \begin{tabular}[c]{@{}c@{}}Lane Straddle\\ Dev\end{tabular} & \begin{tabular}[c]{@{}c@{}}Attack Goal\\ Dev\end{tabular} \\ \midrule
% ka-local31 & 1.39 & 0.7 & 1.3 \\
% ka-local33 & 1.06 & 0.7 & 1.3 \\
% ka-highway36 & 1.18 & 0.7 & 1.9 \\
% ka-highway18 & 1.13 & 0.7 & 1.9 \\ \bottomrule
% \end{tabular}
% \end{table}

% risk to build upon consecutive frames, clearly say we assumed the max attack capability, instant influence on physical world deviation, can precisely control timing and deviationr

% To detection: prevent detection, achieve large deviation Table~\ref{tbl:stealthy_detection}.

% highway18 max adaptive deviation:
% 1.13 m = 0.43 (LD max deviation in benign case) + 0.6 (cusum bias) + 0.1 (cusum threshold)
% highway36
% 1.18 m = 0.48 (LD max deviation in benign case) + 0.6 (cusum bias) + 0.1 (cusum threshold)
% local31
% 1.39 m = 0.59 + 0.7 + 0.1
% local33
% 1.06 m = 0.36 + 0.6 + 0.1

% \junjie{Future work: need an evaluation where we attack from LD side to trigger detection, but no adaptive design in AR.}

\begin{figure}[tbp]
\centering
\includegraphics[width=\columnwidth]{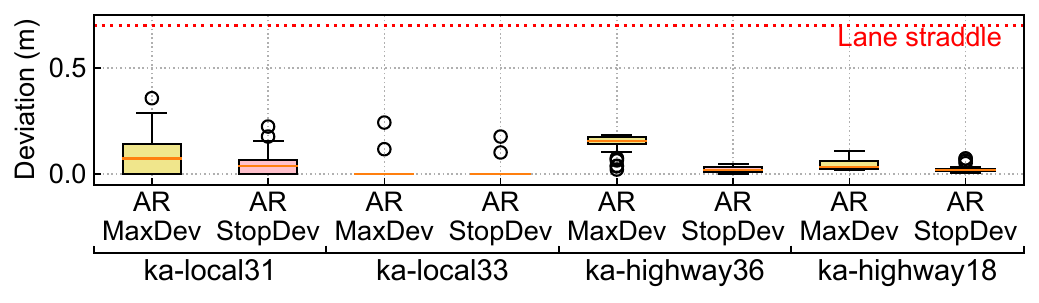}
% \vspace{-0.3in}
\caption{Maximum and stopping deviations during the AR period under the LD attack.}
\label{fig:ld_attack_devs}
% \vspace{-0.05in}
\end{figure}

% \vspace{-0.05in}
\nsubsection{LD-side Adaptive Attack Evaluation} \label{sec:ld_attack}
\vspace{0.05in}

% \newparts{
% Since \ld{} uses lane detection as a defense source, a direct adaptive attack direction is thus LD-side attacks. Therefore, 
% In this section, we evaluate \ld{} against LD-side attacks.}
% , which are direct adaptive attack direction to \ld{}.}

\newparts{
\textbf{Evaluation methodology.}
We explore the defense capability of \ld{} against the latest LD attack in production low-level AD systems, 
named Dirty Road Patch (DRP) attack~\cite{sato2021dirty}, which is designed to affect the detected lane line shapes to mislead the automated lane centering system to drive the vehicle out of the lane boundaries. 
In LD, the lane line shapes are represented as polynomial functions, which are used in \ld{} to calculate the vehicle's lateral deviations  (Appendix~\ref{app:design_impl}). 
% Therefore, in our evaluation, we focus on the attacked lane line polynomials influenced by DRP attack. 
From the 40 attack traces used in the original DRP attack paper, we extract the attacked lane line polynomials in each frame and calculate an averaged LD deviation trace.
In \ld{} design, LD attacks cannot disrupt the driving behaviors before the attack is detected since only MSF outputs are used for navigation at this moment.
To cause vehicle deviations, the LD attack has to trigger the detection in the first place and affect the \textit{fused} localization in the AR period (\S\ref{sec:design_ar}) in order to affect the vehicle control. Therefore, we focus on the AR period in our evaluation. 
To model the DRP attack effect, we apply the deviation trace (start from the detection deviation 0.7 m) to the LD side in the KAIST traces.
Since the MSF side is benign and should generally well-align with the physical positions of the vehicle, we set the MSF outputs in the AR period with the same deviation as the fused localization, but to the opposite direction based on the control assumption (\S\ref{sec:background_threat_model}).}

\newparts{
\textbf{Results.} Fig.~\ref{fig:ld_attack_devs} shows the maximum and stopping deviations in KAIST traces. As shown, none of them is able to even cause lane straddling. On average, the maximum and stopping deviations in the AR period are only 0.08 m ($\delta=0.08$ m) and 0.02 m ($\delta=0.03$ m), respectively. Such a result indicate that \ld{} is quite robust to adaptive attack to the LD side as well. This is because the safety-driven fusion (\S\ref{sec:design_ar}) in \ld{} can effectively penalize the more aggressive source in the driving context, which in this case is the attacked LD outputs, and prevent the fused localization from being influenced by it.}

% \vspace{-0.05in}
\nsection{Limitations Discussion} \label{sec:discuss}
% \vspace{0.05in}

\textbf{Defense coverage of lane detection.}
% As the first work to discover and leverage information sources available in high-level AD systems, we choose to use the lane boundary as a defense-suitable one for attack detection and response against lateral-direction localization attacks. 
In this work, we are the first to explore the novel usage of LD for defense. However, as a defense relying on LD, a potential limitation is the lane line marking coverage. However, as we analyzed in \S\ref{sec:design_overview}, the non-deterministic nature of attacks to MSF localization greatly alleviate such a limitation, where an LD-based defense has the potential to defend against the majority (99.2\%) of the attack attempts. In addition, for important AD applications such as autonomous trucks, they are naturally not subject to such limitation as they mostly operate on highways~\cite{tusimple_ad_truck_middle_mile, walmart_ad_truck_middle_mile}.
% However, the applicability of LD are limited in areas with lane line markings. 
At design level, since high-level AD systems come with semantic maps with accurate road geometry information, \ld{} knows exactly where are the regions without lane line markings and can temporarily disable the defense in such regions (\S\ref{sec:design_detection}).
To address this limitation, a potential future improvement is to also consider other road markings available in such regions, e.g., stop lines~\cite{lin2017stop} and crosswalk markings~\cite{bailo2017robust} in intersections, to help localize the vehicle and to detect MSF deviations. Nevertheless, it is unclear how prevalent such road markings are and how mature and robust the existing perception algorithms are to recognize such road markings.
%\alfred{talk about future improvement direction? if it is a pure argument, it is not necessary to repeat it here.} \junjie{added.}
% a good thing is that as a defense integrated in the AD system, \ld{} knows exactly where it is not applicable and can decide to temporally disable the attack detection (\S\ref{sec:design_detection}).
% In addition, \ld{} can already cover challenging scenarios such as highways, where the lane line markings are generally available, and the attack consequences can be more severe due to the high speeds.
% Nevertheless, a promising extension of this work is to also include information sources that are available in the intersections, e.g., stop lines, to help localize the vehicle and cross-check the MSF deviations.

% \todo{MSF is opportunistic; highway doesn’t have intersections and generally always have lane boundary markings.}

% \todo{mention the FusionRipper intersection attack success rate here: attack success rate:
% 0.83\% = 15 / 1813 (among all attack traces);
% 1.41\% = 15 / 1062 (among all local attack traces)}

% \todo{When talking about the number here, talk from victim’s point of view: Percentage of my driving is protected: 99.17\%.}

% Road markings as an independent source

\textbf{Simultaneous attacks to MSF and LD.}
Since \ld{} leverages LD to detection lateral-direction attack on MSF,
attacks that simultaneously target MSF and LD can thus potentially bypass our detection. 
In fact, such a vulnerability is a general limitation for CPS security research that uses sensor cross-checking/fusion for defense purposes~\cite{feng2018efficient, feng2017efficient, aguilar2017developing, tanil2017detecting, khanafseh2014gps, lee2015gps}. However, in practice, the defense value of \ld{} highly depends on \textit{whether such a simultaneous attack already exists or can be easily achieved}. For MSF and LD, neither of them holds today, since (1) although individual attacks on MSF or LD exist, no existing work shows that they can be effectively \textit{coordinated and synchronized} to achieve simultaneous attack effect control, and (2) it is far from trivial to achieve this with existing individual attack vectors. Specifically, among the attack vectors on camera~\cite{petit2015remote, yan2016can, nassi2020phantom, sayles2021invisible, kohler2021they, sato2021dirty, kang2020lane, jing2021too}, only three works~\cite{nassi2020phantom, sato2021dirty, jing2021too} actually evaluated and shown attack effectiveness on LD in realistic AD settings. All these three works consider adding malicious patterns to the ground (e.g., via road patch or stickers) as the attack vector. However, considering the non-deterministic nature of the existing high-level localization attacks (\S\ref{sec:opportunity}), it would be hard, if not impossible, for the attacker to figure out where to place the attack pattern beforehand, not to mention how to carefully synchronize the malicious pattern with the localization-side attack to effectively bypass \ld{}. Therefore, we consider such simultaneous attack design neither already exists nor can be easily achieved, and leave the systematic exploration of its feasibility as a future direction.

\textbf{Delay between attack and detection.} Another limitation is that our detection and response happen after the attack has occurred to some extent (i.e., some deviations have already been caused by the attack). Even though our system can greatly reduce the safety consequences and transition the vehicle into a minimal-risk condition, it is still better if we can detect the attack immediately after the first injection is sent to the system. We thus consider this as another future direction.

\cut{ % revision: not a limitation any more
\textbf{Lack of closed-loop in real vehicle experiments.} 
Although we include closed-loop evaluations with the AD control and simulation world feedback in the end-to-end simulations (\S\ref{sec:simulation}), the evaluations using the real vehicle (\S\ref{sec:eval_real_car}) still adopt a trace-based evaluation where we model the control effect by assuming the MSF deviations will result into a same amount of physical world deviation in the opposite direction. Ideally, a similar closed-loop evaluation should also be conducted on real AD vehicles to more realistically validate the defense capabilities. However, this is way beyond the affordability of normal academia research groups, and actually even companies such as Waymo and Uber also heavily rely on trace-based and simulation-based evaluations when developing and testing their AD systems for safety and budget considerations~\cite{bansal2018chauffeurnet, frossard2018end}. We thus leave this to future work, e.g., we have an ongoing effort to develop a chassis capable of closed-loop control with AD sensor setup, which might be used for this purpose once built and tested.
}

%However, we currently do not have the necessary hardware and compatible vehicle to run the complete AD system pipeline. 
%We are in the process of developing a real-vehicle sized chassis capable of closed-loop control with AD sensor setup, and plan to use it for validating \ld{} in the future.

%have plans to purchase a AD development chassis with the complete AD sensor set, including GPS, LiDAR, IMU, Camera, etc. We plan to validate \ld{} using this vehicle in a real-world testing facility in the future.

\cut{
\nsubsection{Alternative Design} \label{sec:discuss_alternative_designs}

\textbf{Apply LD as a fusion input in MSF.}
In \ld{}, we use LD outputs in a post-processing step in AD localization to cross validate the MSF lateral deviations. An alternative design might be to directly include LD \textit{as one of the fusion inputs in MSF localization} to mitigate lateral-direction attacks. However, such a design faces several challenges:
(1) it is unclear how to practically fuse relative positioning sources such as LD with global positioning sources such as GPS and LiDAR locator without degrading MSF accuracy. Existing works are able to estimate an LD-based global localizations using semantic maps~\cite{kang2020lane, evlampev2020map}, however, their positioning accurate are at 0.5m level, which might severely degrade the resulting MSF accuracy (at centimeter-level~\cite{wan2018robust}) if got fused together;
(2) additional fusion inputs may be able to increase the robustness against \fr{} to certain degree, but it cannot fundamentally prevent \fr{}~\cite{fusionripper}. On the other hand, \ld{} is able to achieve perfect detection performance against \fr{} and can safely steer the AD vehicle to stop in the ego lane after detection.
}

% \textbf{Attack response after stopping.}
% Call support and police, reboot system, etc. 
% In our emergency trajectory generation (\S\ref{sec:design_ar}), we focus on the most basic requirements, i.e., slowing down and steering. However, real world drivings are more complicated in which the emergency trajectory may stop in the middle of an intersection. (This is still part of the applicable domain issue for lane boundaries). To address, a potential solution is to design a speed profile that consider that can allow the vehicle to pass the intersection before stops.

% should also consider the front obstacles. Generally it is not a concern since the AD planning maintains a distance larger than the stopping distance with the front vehicles in normal driving. However, in the cases such as a vehicle suddenly cuts in and stops, we can do nothing.. To address this, a straightforward solution is to reuse the planning logic in high-level AD systems for 

\cut{
\textbf{Alternative attack response goals.}
In the current \ld{} design, we set the AR goal as to stop with in lane boundaries as soon as possible. Although such an AR goal can prevent more serious consequences that could occur if the vehicle stops out of lane boundaries, e.g., being hit by another vehicle that failed to yield in time, it may be more desirable to pull over to the roadside or even keep driving in the ego lane if possible. However, such AR goals requires more sophisticated AR designs, and thus we leave them as future work.
}

% Instead of stopping in lane. Use the fused EH pose to navigate to pull over to the road side. 
% \vspace{-0.05in}
\nsection{Related Work} \label{sec:related_work}
% \vspace{-0.02in}

\textbf{AD system security.} Prior works have studied attacks and defenses of AD system components for environment sensing and decision-making, such as object detection, tracking, localization, lane detection, and planning~\cite{sp:2021:ningfei:msf-adv, ndss:2022:ziwen:planfuzz, sato2021dirty, arxiv:2022:shen:sok, sato2021wip, sato2020hold, dipalma2021security, ma2023wip, wang2022poster, huai2023doppelganger, jia2020fooling, cao2019adversarial, shin2017illusion, shen2023detecting}. Specifically for attacking localization, sensor spoofing/jamming attacks targeting GPS, LiDAR, IMU, camera, RADAR~\cite{fusionripper, zeng2018all, popperccs11, utaustinspoofer, narain2018security, kerns2014unmanned, cao2019adversarial, petit2015remote, son2015rocking, yan2016can} have been proposed. Only \fr{}~\cite{fusionripper} is able to break the MSF localization on high-level AD systems and cause lateral deviations in MSF outputs. Thus, we target \fr{} and show that \ld{} can effectively detect \fr{} and steer the vehicle to safely stop in the ego lane.

%\textbf{AD localization attacks.}
%Since AD systems rely on sensors for localization, prior works proposed sensor spoofing/jamming attacks targeting GPS, LiDAR, IMU, camera, RADAR~\cite{fusionripper, zeng2018all, popperccs11, utaustinspoofer, narain2018security, kerns2014unmanned, cao2019adversarial, petit2015remote, son2015rocking, yan2016can}. Among them, only \fr{}~\cite{fusionripper} has shown to be able to break the MSF localization on high-level AD systems and cause lateral deviations in MSF outputs. Thus, we target \fr{} and show that \ld{} can effectively detect \fr{} and steer the vehicle to safely stop in ego lane.

% focus on defending against such lateral-direction localization attack.

\textbf{Physical-invariant based defenses.}
Recently, researchers propose physical-invariant based defenses, CI~\cite{ci} and SAVIOR~\cite{ savior}, to detect sensor attacks such as GPS spoofing by cross-checking sensor measurements with system state estimations based on the physical invariants, i.e., the relationships between system states and control inputs. However, as shown in \S\ref{sec:eval_detection}, 
the direct adaptation of existing physical-invariant based approach 
% to the AD context suffers from very high false positives and is actually close to random guessing.
% the detection effectiveness of physical-invariant based detection methods are 
is largely limited because of the complexity of physical dynamics and much smaller attack deviation goals in the AD context.
% mostly because the existing state estimation models, e.g., bicycle model~\cite{kong2015kinematic}, are not accurate enough to detect stealthy attacks such as \fr{}. 
In addition, 
% both works evaluate on robotics systems such as drones and ground rovers, and 
none of them has proposed attack response designs, which is especially important for AD systems (\S\ref{sec:design_overview}). Nevertheless, such physical-invariant based attack detection methods are complimentary to \ld{} and can be incorporated into our design for attack detection if the accuracy of state-estimation model can be further improved.

\textbf{Attack response/recovery.}
According to a survey on the broader Cyber-Physical Systems security, existing defenses mostly focus on attack detection and very few works studied attack responses~\cite{giraldo2017security}. Particularly, Choi et al.~\cite{choi2020software} and Zhang et al.~\cite{zhang2020real} recently propose \textit{attack recovery} methods, which apply similar state estimations as above to replace attacked sensors in the attack recovery period. Thus, they suffer from the same model accuracy limitations in the AD context.
Moreover, they intend to maintain normal operations of the system for a short duration until the system is taken-over by the human driver, which does not exist on high-level AD vehicles when deployed commercially~\cite{baidu_driverless_robotaxi, waymo_driverless}.
Additionally, attack responses in high-level AD systems require more careful design on AR trajectories (\S\ref{sec:design_ar}) to safely navigate the vehicle.
\nsection{Conclusion} \label{sec:conclusion}
% \vspace{-0.02in}

In this work, we perform the first systematic exploration of the novel usage of lane detection (LD) to defend against lateral-direction attacks in high-level AD localization. We design the first domain-specific LD-based defense, \ld{}, that is capable of both real-time attack \textit{detection} and \textit{response}. Our evaluation on real-world AD sensor traces show that \ld{} is much more effective than directly-adapted physical-invariant based defenses at attack detection with accurate and timely detection. 
We also show that \ld{} can safely stop the vehicle in the current lane upon detection. We implement \ld{} on two open-source high-level AD systems and evaluate its effectiveness under end-to-end driving with closed-loop control in both simulation and the physical world. We also evaluate against two adaptive attacks and find that \ld{} is robust to an idealized stealthy attack that aims to evade detection and the latest LD-side attack targeting the response stage.
\section*{Acknowledgment}

We would like to thank the anonymous reviewers for their valuable feedback on our work. This research was supported in part by the NSF under grants CNS-1929771, CNS-2145493, and USDOT under grant 69A3552047138.
%\newpage

\bibliographystyle{ieeetr}
{
\footnotesize
\bibliography{ref_ld3, ref_fusionripper, ref_msf_survey, ref_reviews}
}

\appendix
% \section*{Appendix}

\begin{figure}[bp]
\centering
\includegraphics[width=.9\linewidth]{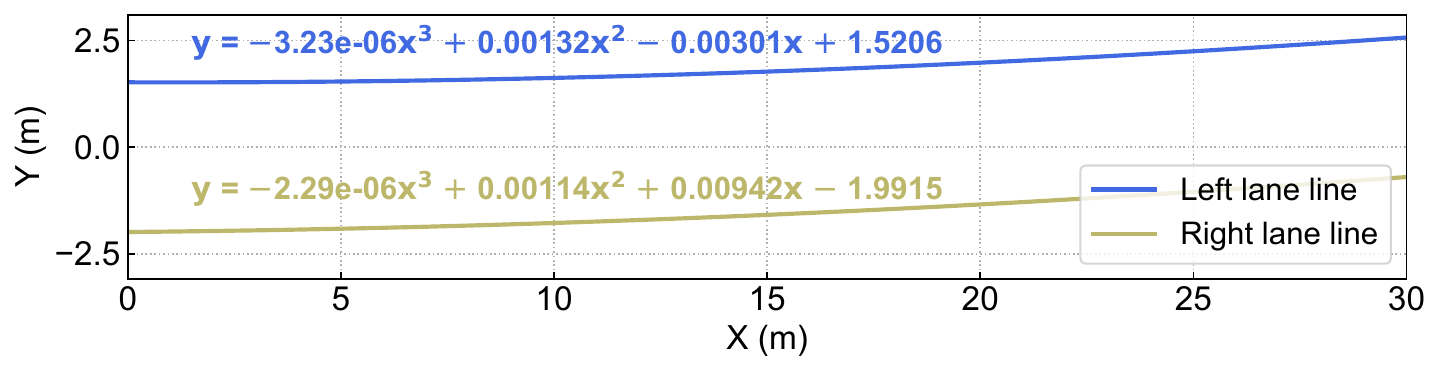}
% \vspace{-0.15in}
\caption{Example of left/right lane line polynomial functions.}
\label{fig:ld_polys}
\end{figure}

% \subsection{Detailed \ld{} Design and Implementation Choices} \label{app:design_impl}
\subsection{Converting LD outputs to lateral deviations} \label{app:design_impl}

% The \ld{} is designed to be generally applicable for high-level AD systems. Specifically, we implement \ld{} on two popular open-source high-level AD systems, Baidu Apollo~\cite{apollo} and Autoware~\cite{autoware}, following the system diagram in Fig.~\ref{fig:design_overview}. 
% In this section, we describe the implementation details in \ld{}.

% \textbf{Converting LD outputs to lateral deviations.}
% For high-level AD systems, although they do not use LD for localization purpose as mentioned in \S\ref{sec:opportunity}, they often still perform LD in the perception module for camera calibration~\cite{apollo}. Specifically, 
The LD output consists of the detected left and right lane lines, which are represented as polynomial functions in the bird's eye view~\cite{apollo, openpilot}. An example of the polynomial functions is shown in Fig.~\ref{fig:ld_polys}. For these polynomial functions, the absolute values at $x=0$ represent the vehicle's distances to the lane lines, $d_{\text{left}}$ and $d_{\text{right}}$. 
Therefore, we can calculate the lateral deviation to the lane centerline by
\vspace{-0.07in}
\begin{equation}
\small
\begin{aligned}
lw/2 - d_{\text{left}}\ \ \text{or}\ \  d_{\text{right}} - lw/2,
\end{aligned}
\label{eqt:lateral_dev_calc}
\vspace{-0.07in}
\end{equation}
where $lw$ is the lane width. 
We calculate the lateral deviation as a \textit{signed} number to differentiate the deviations to the left (positive) and to the right (negative). 
%\alfred{why not use absolute} \junjie{re-wrote the equation and explained why we use a signed lateral deviation.}

Although we can also obtain the lane width from the lane line polynomials (i.e., $lw_{\text{poly}} = d_{\text{left}} + d_{\text{right}}$), as mentioned in \S\ref{sec:design_detection}, it is not uncommon that one of the lane lines is missing or incorrectly detected in real world driving, e.g., when the current lane splits into a through lane and a left or right turn lane.
In such cases, directly using the distances from the polynomial functions would result into a wrong lateral deviation. To address this, we include two optimizations in the lateral deviation calculation: (1) instead of estimating the lane width from the polynomial functions, we query the current lane width from the semantic map (line 4 in Alg.~\ref{alg:attack_detection}), and (2) prioritize the lane line with a smaller distance to the vehicle by using it to calculate the lateral deviation in Eq.~\ref{eqt:lateral_dev_calc} (line 4, 6 in Alg.~\ref{alg:ld_dev_calc}).
%\alfred{how to ``prioritize''? the equation above needs input from both left and right lane line poly.}\junjie{how about the current version?} 
This is because for the lane splitting scenario mentioned above, the incorrectly-detected lane line often has a much larger distance compared to the correctly-detected one.
A special handling is that when both lane lines are incorrectly detected, which is very rare in SCNN~\cite{pan2018spatial} and never occur in OpenPilot LD model~\cite{openpilot}, we will reuse the previously calculated lateral deviation. 
% Note that this is different from the cases where both lane lines are naturally unavailable, e.g., in intersections, where we disable the attack detection since they are out of the applicable domain for lane detection (\S\ref{sec:design_detection}).

\begin{algorithm}[tbp]
\footnotesize
\caption{Calculation of LD deviation to lane centerline}
\label{alg:ld_dev_calc}
\textbf{Notations:} $lw_{\text{map}}$: lane width from map; $poly(\cdot)$: polynomial function fitted on detected lane line; $d$: distance to lane line; $D$: deviation to lane centerline
\begin{algorithmic}[1]
\Function{\text{L\textsc{d}}\text{D\textsc{ev}}}{$LD$, $lw_{\text{map}}$}
    \State $d_{\text{left}} \gets \lvert LD.poly_{\text{left}}(0) \rvert$ \textbf{if} $LD.poly_{\text{left}}$ \textbf{else} $\infty$ \Comment{dist. to left line}
    \State $d_{\text{right}} \gets \lvert LD.poly_{\text{right}}(0) \rvert$ \textbf{if} $LD.poly_{\text{right}}$ \textbf{else} $\infty$ \Comment{dist. to right line}
    \If{$LD.poly_{\text{left}}$ \textbf{and} $d_{\text{left}} < d_{\text{right}}$}  \Comment{left line is correct}
        \State $D$ $\gets$ $lw_{\text{map}}/2 - d_{\text{left}}$ \Comment{dev. to centerline; $+$: left, $-$: right}
    \ElsIf{$LD.poly_{\text{right}}$ \textbf{and} $d_{\text{right}} < d_{\text{left}}$} \Comment{right line is correct}
        \State $D$ $\gets$ $d_{\text{right}} - lw_{\text{map}}/2$
    \Else \Comment{if none of the lane lines are correctly detected}
        \State $D$ $\gets$ last calculated $D$ \Comment{re-use last dev. to centerline}
    \EndIf
    \State \textbf{return} $D$
\EndFunction
\end{algorithmic}
\end{algorithm}

% \textbf{Safe deceleration in attack response.}
% Generally, a deceleration $<$4.6 $\mathrm{m/s^2}$ is considered as safe for maintaining steady control~\cite{deceleration}. Thus, to calculate the speed profile of the AR trajectory (\S\ref{sec:design_ar}), we apply 4 $\mathrm{m/s^2}$ as the deceleration, which is also defined in Baidu Apollo as the maximum allowed deceleration to ensure safety~\cite{apollo}. 

% \vspace{-0.1in}
\subsection{Evaluation of LiDAR Localization Dependency on Lane Line Markings}  \label{app:lidar_lane_line_dependency}
% \vspace{-0.1in}

\textbf{Evaluation methodology.} 
To evaluate the dependency of LiDAR localization on lane line markings, we first create two traces of modified LiDAR data: one without lane line markings (denote as \textit{no-marking}) and another with incorrect lane line markings (denote as \textit{wrong-marking}). Next, we execute the LiDAR locators on the original LiDAR trace as well as on the two modified traces. \textit{If a LiDAR locator does not rely on the lane line markings, we should observe a high similarity between the original and the modified executions.}

Specifically, LiDARs scan the surrounding environment and output Point Cloud Data (PCD), which stores the 3D positions and intensities of the reflected laser points. Since the lane line markings will exhibit distinctively higher intensities than the other road surface due to their color differences, we create the \textit{no-marking} PCDs by changing their intensities to the same as other road surface area. 
To do that, we first apply the commonly-used RANSAC plane segmentation~\cite{schnabel2007efficient} on the PCDs to find all points that belong to the ground plane, i.e., the road surface, and then set the intensities of these ground points to their median value. This thus effectively makes the lane line markings indistinguishable from the other road surface.
The creation of \textit{wrong-marking} PCDs is slightly more complicated. After recognizing all ground points, we identify the lane line marking points depending on whether their intensities are above a certain threshold. For each lane line marking point, we search a corresponding ground point that is \textit{laterally offset by half-lane-width} and set their intensities the same as the original lane line marking points. Finally, we clear the original lane line marking points by setting their intensities to the median ground point intensities. Since the lane line markings are moved by half-lane-width, the \textit{wrong-marking} PCDs should have the largest lateral LiDAR localization impact if the lane line markings have any effect on the LiDAR locator.
Fig.~\ref{fig:laneline_removal_example} shows such an example of the original PCD and the one with \textit{no-marking} and \textit{wrong-marking}. 
% Note that the LiDAR map used in the LiDAR locator still have the complete lane line markings.

\textbf{Experimental setup.} 
We evaluate on 2 LiDAR locators, one from Baidu Apollo (BA-LiDAR locator)~\cite{wan2018robust} and another from Autoware (AW-LiDAR locator)~\cite{autoware}. Details of the LiDAR locators can be found in Appendix~\ref{app:lidar_locators}. Since MSF localization takes not only position measurements but also position uncertainties from LiDAR locator as inputs, we calculate both the \textit{position accuracies} and \textit{uncertainty correlation} with the original and no/wrong-marking PCDs to show the similarity.
% with and without lane line markings removal to show the similarity.
We evaluated on the same 5 local road and highway traces in \fr{}~\cite{fusionripper} from two datasets. For each trace, we exclude intersections since they do not have lane line markings. Among them, since \textit{ba-local} does not provide ground truth positions, we calculate the position accuracy based on the LiDAR locator with the original lane line markings.

\textbf{Results.}
Table~\ref{tbl:correlation} shows the experiment results. For the position accuracy, we report the Root Mean Squared Error (RMSE) between the LiDAR locator positions and the ground truth positions or the ones with the original lane line markings. For the correlation, we use the commonly-used Pearson's correlation, and a correlation coefficient $>$0.5 is considered strongly correlated~\cite{cohen2013statistical}. As shown, for both LiDAR locators, the uncertainty correlation coefficients between the original and modified PCDs are all well above the threshold for strong correlation, and their position accuracies are also all at centimeter-level. Particularly, since AW-LiDAR locator does not use lane line markings at the design level (Appendix~\ref{app:lidar_locators}), the traces consequently show perfect correlations and identical position accuracies no matter how we modify the lane line markings. Such a result suggests that the existing LiDAR locators used in high-level AD systems are indeed largely ignore the lane line marking information when localizing the vehicle on the map, which might because global localization focuses more on the unique features on the road, such as buildings, roadside layouts, and traffic signs. 
As a result, this indicates that lane line markings are largely independent of the ones that are already used in high-level AD localization and thus pose a great potential for defense purposes.

\begin{table*}[tbp]
\footnotesize
\begin{minipage}{0.23\linewidth}
	\centering
    \includegraphics[width=.9\columnwidth]{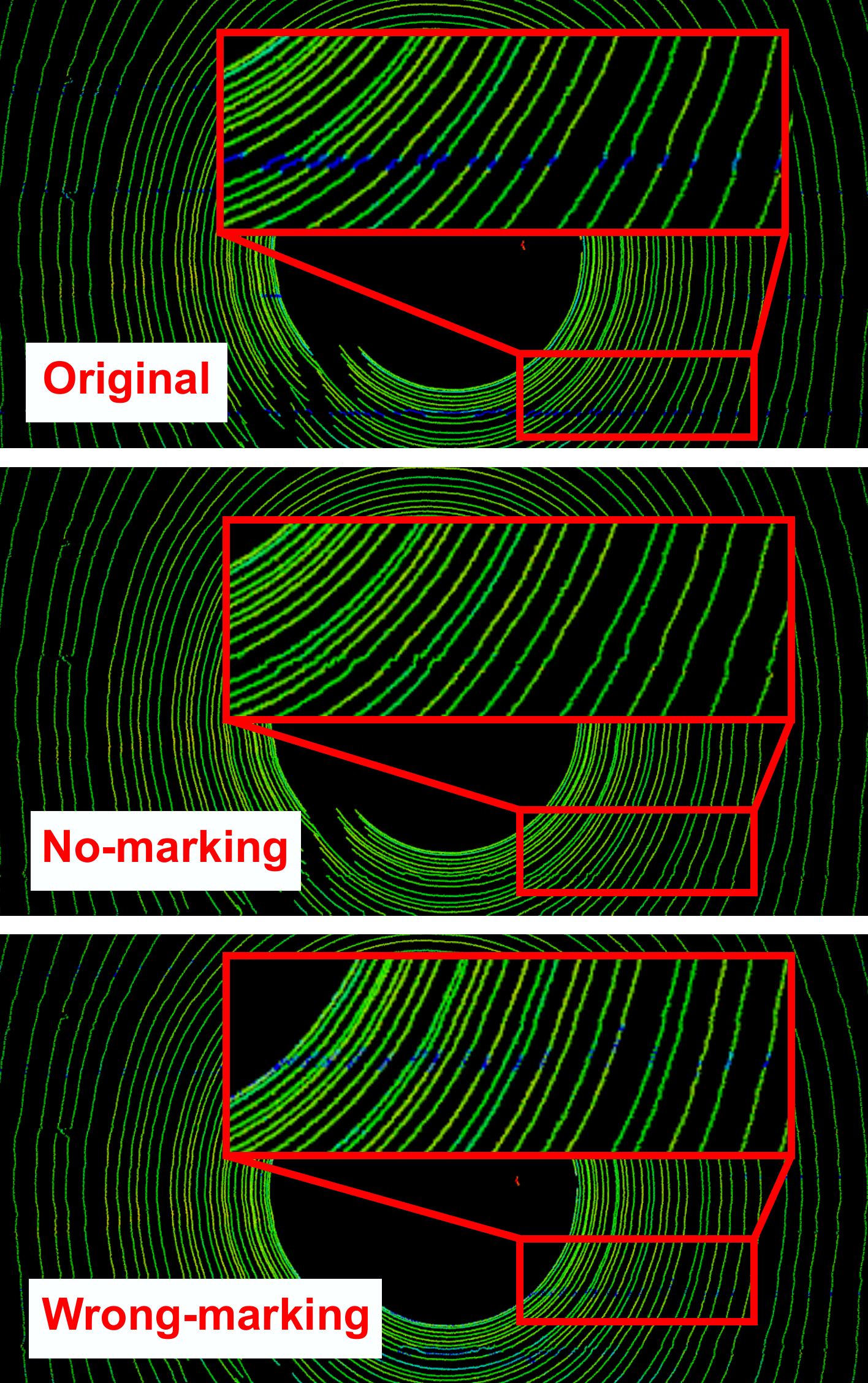}
    % \vspace{-0.05in}
    \captionof{figure}{PCDs with the original, removed, and incorrect lane line markings.}
    \label{fig:laneline_removal_example}
\end{minipage}\hfill
\begin{minipage}{0.76\linewidth}
    \centering
    \caption{The uncertainty correlation coefficients ($r$) and position accuracies (RMSE) of LiDAR locators using the original, lane line markings removed (denote as \textit{no-marking}), and incorrect lane line markings PCDs (denote as \textit{wrong-marking}). Results with statistically strong correlation are highlighted in \textbf{bold}; we omit the $p$-values as they are all statistically significant. The numbers are averaged across all sensor traces used in \fr{}~\cite{fusionripper}.}
    \label{tbl:correlation}
    % \vspace{-0.1in}
    % \setlength{\tabcolsep}{5.5pt}
    \begin{tabular}{@{}c|cc|ccc@{}}
    \toprule
     & \multicolumn{2}{c|}{Uncertainty Correlation ($r$)} & \multicolumn{3}{c}{Position Accuracy (RMSE)} \\ \cmidrule(l){2-6} 
     & \begin{tabular}[c]{@{}c@{}}Original vs\\ No-marking\end{tabular} & \begin{tabular}[c]{@{}c@{}}Original vs\\ Wrong-marking\end{tabular} & Original & No-marking & Wrong-marking \\ \midrule
    BA-LiDAR Locator & \textbf{0.89} & \textbf{0.64} & 0.064 m & 0.065 m & 0.063 m \\
    AW-LiDAR Locator & \textbf{1.0} & \textbf{1.0} & 0.076 m & 0.076 m & 0.076 m \\ \bottomrule
    \end{tabular}

    % \vspace{-0.1in}
    \caption{Semantic map APIs required for \ld{}.}
    \label{tbl:map_apis}
    % \vspace{-0.1in}
    % \setlength{\tabcolsep}{3pt}
    \begin{tabular}{@{}ll@{}}
    \toprule
    Map API & Description \\ \midrule
    \texttt{MapLaneDev(pose)} & Query the deviation from \texttt{pose} to the closest lane centerline \\
    \texttt{MapLaneWidth(pose)} & Query the width of the closest lane to \texttt{pose} \\
    \texttt{MapLanePoint(pose)} & Query the closest point and lane heading on the closest lane centerline to \texttt{pose} \\
    \texttt{MapIsIntersection(pose)} & Query if \texttt{pose} is located in an intersection\\ \bottomrule
    \end{tabular}
\end{minipage}
\end{table*}

\subsection{Details of the LiDAR Locators} \label{app:lidar_locators}

At design level, Baidu Apollo LiDAR locator (BA-LiDAR locator)~\cite{wan2018robust} considers point cloud intensities in its position calculation. Thus, the modifications of lane line intensities do have the potential to affect the BA-LiDAR locator performance. On the other hand, Autoware LiDAR locator (AW-LiDAR locator)~\cite{autoware} only uses the position data in the PCD and completely ignores the intensities. This means that AW-LiDAR locator does not consider lane line markings at the design level.
% Nevertheless, we still include AW-LiDAR locator in our evaluation.
% When used in AD localization, the MSF takes not only the position measurement but also the position uncertainty from LiDAR locator as inputs. Thus, we report the \textit{position accuracies} and \textit{uncertainty correlation} with and without lane line markings removal to show the similarity. 
Since AW-LiDAR locator implements the Normal Distributions Transform (NDT) algorithm~\cite{ndt}, which does not output position uncertainty by default, thus we follow a common adaptation for NDT to use the point cloud matching fitness score as the uncertainty~\cite{merten2008three}.

% We evaluated on the same 5 local road and highway traces in \fr{}~\cite{fusionripper} from two datasets. For each trace, we exclude the road segments that do not have lane line markings, e.g., intersections, since they are out of the applicable domain for lane boundaries. Among them, since \textit{ba-local} does not provide ground truth positions, we calculate the position accuracy based on the LiDAR locator with the complete lane line markings instead of the ground truth positions.

\subsection{SAVIOR Evaluation Setup} \label{app:savior_setup}

To evaluate SAVIOR, we follow the similar methodology as the ground rover evaluation in the SAVIOR paper~\cite{savior}, i.e., using the kinematic bicycle model~\cite{kong2015kinematic} and an Extended Kalman Filter (EKF) to predict the system state (i.e., position in x, y coordinates) given the vehicle control commands (i.e., steering and acceleration). Although the vanilla bicycle model does not have tunable parameters, we follow a similar implementation as SAVIOR by adding coefficients to the bicycle model equations~\cite{savior_code}. Same as SAVIOR, we use the \textit{nlgreyest} system identification tool from Matlab~\cite{matlab_si} to find the coefficients that can best fit the sensor and control trace.
During the evaluation, we continuously calculate the residuals between the GPS measurements and the predicted positions from the EKF, and feed the residuals to a CUSUM anomaly detector for attack detection. An execution that triggers the CUSUM detector will be considered as under attack.

Since the KAIST dataset~\cite{jeong2019complex} does not store the control commands when the traces were collected, which are required for the SAVIOR evaluation, we replay the KAIST sensor traces as inputs to Baidu Apollo v5.0.0~\cite{apollo} to collect the control module outputs, i.e., steering and throttle commands. In particular, the control module calculates such commands based on the localization and a planned trajectory, which is a sequence of trajectory points that the vehicle should follow. However, the planned trajectory is runtime information optimized by the planning module during driving, which is not available in the dataset. Since the ground truth positions in the KAIST traces represent the trajectory points followed by the AD vehicle, we thus convert the ground truth positions into planned trajectories according to the format in Baidu Apollo and use them as one of the control inputs. With the planned trajectories, we then feed the benign localization and attacked localization outputs to obtain the benign control commands and attack influenced control commands respectively.

In addition to the KAIST traces, we also evaluate SAVIOR on a dataset that contains the original control commands to validate SAVIOR's detection performance in an ideal setting. However, similar performance is observed in that dataset to the ones on KAIST traces. More details of this are in Appendix~\ref{app:savior_with_real_control}.

\subsection{Evaluating SAVIOR on Dataset with Control Commands} \label{app:savior_with_real_control}

\begin{figure}[h]
\centering
\includegraphics[width=0.65\linewidth]{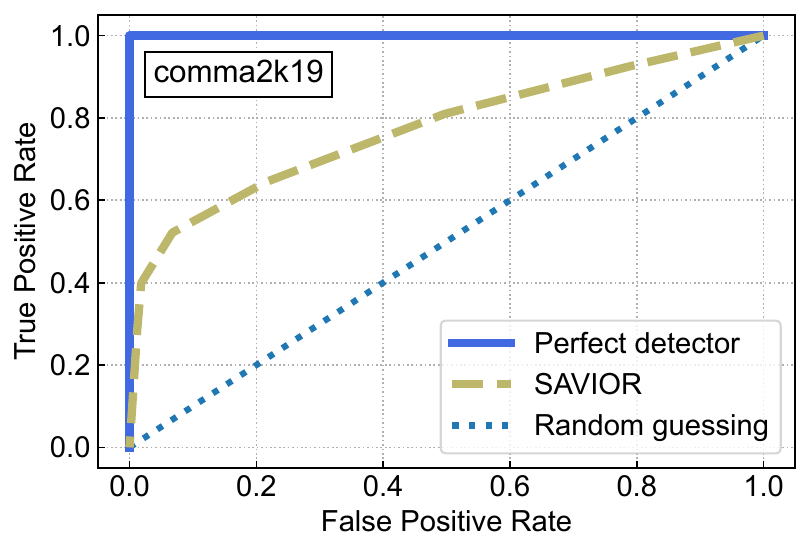}
% \vspace{-0.1in}
\caption{Attack detection ROC curve of SAVIOR on comma2k19~\cite{comma2k19} with original vehicle control commands.}
\label{fig:savior_comma2k19}
% \vspace{-0.2in}
\end{figure}

Since SAVIOR requires vehicle control commands, for which we collected by replaying the KAIST traces in Baidu Apollo in our evaluation (Appendix~\ref{app:savior_setup}), one might argue that SAVIOR may perform much better if given the originally collected vehicle control commands. Therefore, we evaluate SAVIOR on the comma2k19 dataset~\cite{comma2k19}, which contains the original vehicle control commands when the traces were collected. Since the comma2k19 dataset does not provide LiDAR data, we thus cannot run the MSF attack. To evaluate SAVIOR, we apply the most aggressive GPS spoofing parameters in the MSF attack ($d=2.0$, $f=2.0$) to the GPS data and examine SAVIOR's capability at detecting such obvious GPS spoofing attempts. As shown in Fig.~\ref{fig:savior_comma2k19}, SAVIOR's detection performance is close to the one on the \textit{ka-highway36} (Fig.~\ref{fig:trace_rocs_devs}) and is still far from a perfect detector.

\subsection{Semantic Map APIs Required in \ld{}} \label{app:map_apis}

As mentioned in \S\ref{sec:design_detection}, the \ld{} design queries the semantic map in high-level AD systems to obtain the lateral deviation to lane centerline or lane width at specific positions. Table~\ref{tbl:map_apis} lists the map APIs for \ld{} and their descriptions. Note that all these map APIs are available in typical high-level AD systems, e.g., Baidu Apollo~\cite{apollo} and Autoware~\cite{autoware}.

\end{document}